\documentclass[superscriptaddress,prd,showpacs,twocolumn,nofootinbib]{revtex4-1}

\usepackage{graphicx}
\usepackage{color}
\usepackage{epsf}
\usepackage{bm}
\usepackage{amsmath}
\usepackage{amsfonts}
\usepackage{amssymb}
\usepackage{epstopdf}%
\newcommand{\lsim}   {\mathrel{\mathop{\kern 0pt \rlap
  {\raise.2ex\hbox{$<$}}}
  \lower.9ex\hbox{\kern-.190em $\sim$}}}
\newcommand{\gsim}   {\mathrel{\mathop{\kern 0pt \rlap
  {\raise.2ex\hbox{$>$}}}
  \lower.9ex\hbox{\kern-.190em $\sim$}}}
\newcommand{\bw}{\begin{widetext}\begin{equation}}
\newcommand{\ew}{\end{equation}\end{widetext}}
\newcommand{\be}{\begin{equation}}
\newcommand{\ee}{\end{equation}}
\newcommand{\bea}{\begin{eqnarray}}
\newcommand{\eea}{\end{eqnarray}}

\begin{document}

\title{Bose-Einstein Condensate dark matter models in the presence of baryonic matter and random confining potentials}

\author{Tiberiu Harko}
\email{tiberiu.harko@aira.astro.ro}
\affiliation{Department of Theoretical Physics, National Institute of Physics
and Nuclear Engineering (IFIN-HH), Bucharest, 077125 Romania,}
\affiliation{Astronomical Observatory, 19 Ciresilor Street, 400487 Cluj-Napoca, Romania,}
\affiliation{Department of Physics, Babes-Bolyai University, 1 Kogalniceanu Street,
400084 Cluj-Napoca, Romania,}
\author{Eniko J. Madarassy}
\email{eniko.madarassy@physics.uu.se}
\affiliation{Department of Physics, Babes-Bolyai University, 1 Kogalniceanu Street,
400084 Cluj-Napoca, Romania,}

\date{\today}

\begin{abstract}
We consider the effects of an uncorrelated random potential on the properties of Bose-Einstein Condensate (BEC) dark matter halos, which acts as a source of disorder, and which is added as a new term in the Gross-Pitaevskii equation, describing the properties of the halo. By using the hydrodynamic representation we derive the basic equation describing the density distribution of the galactic dark matter halo, by also taking into account the effects of the baryonic matter, and of the rotation. The density, mass and tangential velocity profiles are obtained exactly in spherical symmetry by considering a simple exponential density profile for the baryonic matter, and a Gaussian type disorder potential.  To test the theoretical model we compare its predictions with a set of 39 galaxies from the Spitzer Photometry \& Accurate Rotation Curves (SPARC) database.  We obtain estimates of the relevant astrophysical parameters of the dark matter dominated galaxies, including the baryonic matter properties, and the parameters of the random potential. The BEC model in the presence of baryonic matter and a random confining potential gives a good statistical description of the SPARC data. The presence of the condensate dark matter could also provide a solution for the core/cusp problem.
\end{abstract}

\pacs{04.20.Cv, 04.50.Kd, 98.80.-k}

\maketitle
\tableofcontents

\section{Introduction}

After almost 100 years of intensive research, the dark matter is still unsolved in a satisfactory way.  Originating from the analysis of Zwicky \cite{Z1,Z2}, who applied the virial theorem to the Coma Cluster and obtained evidence of the presence of a large amount of unseen mass,   whose gravitational attraction must hold the cluster together, a large number of cosmological observations have confirmed that such a mysterious component must be present in the Universe. In particular, the Planck satellite data, resulting from the final full-mission Planck measurements of the cosmic microwave background (CMB) anisotropies,  combining information from the temperature and polarization maps and the lensing reconstruction, indicate a dark matter density $\Omega _ch^2 = 0.120 \pm 0.001$, while the baryon density is only of the order of  $\Omega _bh^2 = 0.0224 \pm 0.0001$  \cite{P}. Thus these data confirm the $\Lambda$CDM standard cosmological scenario, in which the dynamics of the Universe is dominated by two essentially unknown components, dark energy and dark  matter, respectively. Hence the estimations of the cosmological parameters from the Planck data on the Cosmic Microwave Background radiation did conclusively indicate that the Universe contains of 74\% dark energy, 22\% non-baryonic dark matter, and about 4\% baryonic matter only \cite{P}.

The rotation curves of most of the observed  galaxies do not follow the simple predictions based on Newton's gravity, or even general relativity. Instead, the rotational velocity curves, as functions of the radial distance $r$ from the galaxy center, remain flat for a large range of radii. This behavior clearly contradicts the predictions of the Newtonian physics, in which the equality of the centrifugal and gravitational force,  $mv^2/r=GmM/r^2$, where $m$ is the particle mass, $M$ the mass of the central object, $G$ the gravitational constant, and $v$ the rotational velocity, implies that $v^2=GM/r$, that is, a decreasing velocity as a function of the radial distance, a behavior which is not observed. The standard solution of this problem is given by assuming the existence around each galaxy of a dark matter halo, in which most of the mass of the galaxy is contained. The dark matter halo must be formed of  matter  composed of non-baryonic matter, which is assumed to be collisionless, and not, or very weakly, interacting with ordinary baryonic matter. For recent reviews on the different aspect of dark matter theories, and the observational status for its search, see \cite{Rev1,Rev2,Rev3,Rev4, Rev5}.

Despite the intense effort made to directly detect dark matter in Earth based or space experiments, up to now to positive outcome has been reported. An important avenue for the detection of the presence of dark matter could be provided by the observation of its
possible decay into ordinary particles. If such annihilation processes do really occur, then the detection of a
large number of particles generated by the dark matter could give some important observational signatures for the presence of dark matter. For example, the AMS-02 experiment on-board the International Space Station has detected several tens of thousands of antiproton events \cite{AMS}. If these results are interpreted as evidence for dark matter, they may indicate a particle of mass $m_{\chi} = 50-100$ GeV, with an annihilation cross section into hadronic final states having an approximate thermal cross section, $\left<\sigma v\right>\approx 10^{-26}$ cm$^2$/s.  However, in \cite{Pr} it was found
that the global significance of the antiproton excess is reduced to below 1$\sigma$ once all systematics are fully taken into account, and in the  mass range 10–10000 GeV no significant evidence for a dark-matter signal in the AMS-02 antiproton data does exist.
The detection of the excess gamma-ray and positron
emissions in our galaxy could also be an indicator for the presence of dark matter, since these signals could be interpreted as coming from the annihilation of dark matter with mass in the range of $m\sim 10 - 100$ GeV. For a detailed discussion of
this problem see \cite{Chan1,Chan2,Chan3,Chan4,Chan5}, where alternative possibilities for the interpretation
of the observational data are also analyzed.

A very important, {\it but still gravitational}, observational evidences for {\it the existence of dark matter} is provided by the observations of a galactic
cluster called the Bullet Cluster  (1E 0657-56), which is formed from two colliding clusters of galaxies. The cluster pair consists  of stars, intra cluster gas, and the dark matter. These components have a different dynamics during the collision, and consequently they can be studied independently through gravitational lensing. Hence, due to the past collision of its two components, in the Bullet Cluster cluster the dark matter and the baryonic matter components are disconnected \cite{Bul}, a result consistent with the prediction that {\it dark matter is only gravitationally interacting with other forms of matter}.

Hence, intensive experimental and observational searches did not succeed up to now to bring some {\it non-gravitational} insights into the dark matter problem. There are still three fundamental questions that wait to be solved, namely, {\it if dark matter really exists}, or it is just a modification of the gravitational force at galactic scales, what are {\it the nature of the dark matter particle, and its mass}, if it exists, and what is {\it the state in which this particle does exist} in the cosmic space.

It is tempting to interpret the observational data without adopting the hypotheses of the existence of dark matter. In this case one must assume that on galactic or extra-galactic scales the laws of gravity (Newtonian or
general relativistic) are modified, and a new fundamental theory describes the gravitational phenomena. One of the first attempts to explain the rotation curves by modifying the laws of Newtonian physics is the MOND (Modified Newtonian Dynamics) theory, initially introduced in \cite{M1}. Alternative theories of gravity have been also extensively used as  possible explanations of the dark matter related observations \cite{M2,M3,M4,M5,M5a, M6,M7,M8,M9,M10,M11,M12,M13,M14,M15}. For example, in \cite{M5a} it was found that  a modification of the Einstein-Hilbert Lagrangian of the form $R^{1+\delta}$, $\delta <<1$, can explain the galactic dynamics without the need of introducing dark matter. For a detailed review of the dark matter problem in some modified theories of gravity see \cite{M16}.

A central problem in the dark matter studies is the nature of the dark matter particle, and of its mass. Theoretical dark matter models can be classified generally into three major types, cold, warm and hot dark matter models, respectively, depending on the energy of the particles composing
it \cite{DM1}. In the cold dark matter scenario, the growth of structure is hierarchical, with astrophysical objects collapsing first under their self-gravity,  and forming later larger and more massive objects. In the hot dark matter model, the cosmological structures do not form hierarchically, and instead they forms by fragmentation, with large superclusters forming first, which then separate into smaller components. The warm dark matter scenario  has properties intermediate between those of cold dark matter and hot dark matter models, respectively. Many types of particles have been proposed as potential candidates for the dark matter, with the two main candidates being the WIMPs (Weakly Interacting Massive Particles) and the axions, respectively \cite{DM1}.
WIMPs are assumed to be heavy particles that interacts through the weak force \cite{DM2,DM3}, and they are not detected yet.
The axions are bosons that were first introduced in elementary particle physics as a possible solution of the strong CP
(Charge+Parity) problem, which consists of quantum chromodynamics not breaking the CP symmetry \cite{DM4,DM5}. As for the proposed mass range of the dark matter particles, it covers a huge spectrum between $10^{-22}$ meV for axions, and around 10,000 GeV for massive WIMPs \cite{DM1}, indicating the extreme uncertainty in the present day knowledge of the properties of dark matter.

The third fundamental property of dark matter is related to the physical state it does exist. This question can be answered  precisely, firmly,  and with a high level of certainty from a physical point of view, since {\it if dark matter consists of bosons, and it is cold}, it must exist in {\it the form of a Bose-Einstein Condensate}. Bose-Einstein Condensation is a well known physical phenomenon, first described by Bose \cite{Bo1}, and by Einstein \cite{Bo2,Bo3}. When in a bosonic gas the temperature becomes smaller than a critical value, all integer spin particles
condense into the lowest quantum state. The essential feature of the condensate phase is that {\it microscopic quantum phenomena} become detectable {\it at the macroscopic level}. From a physical point of view the necessary condition for the Bose-Einstein Condensation process to occur is that the particles in the gas must become correlated quantum mechanically. This condition is satisfied when the de Broglie thermal wavelength $\lambda _T$ becomes  greater than the mean interparticle distance $d$. The phase transition of the boson gas to the condensate state of the boson gas begins when the temperature
$T$ becomes smaller than the critical temperature $T_c$, given by \cite{Nat1, Nat2,re1, re2, re3, re4},
\begin{equation}\label{Ttr}
T_{cr}=\frac{2\pi\hbar^2\rho_{cr}^{2/3}}{ \zeta^{2/3}(3/2)m^{5/3}k_B},
\end{equation}
where by $\rho_{cr}$ we have denoted the critical transition density, $m$ is the mass of the particle in the boson gas, $k_B$ is Boltzmann's constant, while $\zeta (3/2)$ is the Riemann zeta function $\zeta (s)=\sum _{n=1}^{\infty}{1/n^s}$, respectively. The experimental realization of
Bose-Einstein Condensates in laboratory represented a major success in condensed matter physics, and was first realized in 1995 by cooling a dilute vapor of rubidium-87 atoms, consisting of around two thousand particles,  to temperatures below 170 nK. In order to cool down the system at such low temperatures a combination of laser and magnetic evaporative cooling techniques were used \cite{exp1,exp2,exp3}. From a physical and experimental  point of view, the appearance in a bosonic system of a Bose-Einstein Condensate is indicated by the presence of sharp peaks in both coordinate and momentum space distributions of the particles \cite{Nat1, Nat2,re1,re2,re3,re4}.

The only experimental/obervational evidence for the presence of Bose-Einstein Condensates on a microphysical scale appeared in laboratory
investigations, which have been done on {\it a very small scale}. However, {\it one cannot exclude a priori the possibility of
the existence of some forms of bosonic condensates outside the Earth}. In high density general relativistic environments, like, for example, quark
or neutron stars, the quarks or the neutrons can form Cooper pairs.  The coupled neutrons or quarks then acquire bosonic properties, and, once the temperature or density attain their critical values, they condense into the ground state, forming the so-called Bose-Einstein Condensate stars. This kind of cosmic object  may have maximum central densities of the order of $%
0.1-0.3\times 10^{16}$ g/cm$^3$, and minimum radii in the range of 10-20 km.  Their maximum masses can reach values of the
order of  $2M_{\odot}$. Detailed investigations of the compact objects consisting of a Bose-Einstein Condensate have been presented in \cite{starsm1, stars0, stars1,stars2,stars3,stars4,stars5,stars6,stars7, stars8, stars9, stars10, stars11}.

According to the present dominating $\Lambda$CDM cosmological paradigm, the second most important component of the Universe is cold dark matter. If one assumes that cold dark matter is present in the form of a {\it bosonic gas}, and if we adopt the basic postulate of the universality of the physical laws, meaning that the same physics govern the behavior of matter on Earth and in the outer space, the possibility that a Bose-Einstein Condensation process may have occurred during the cosmological evolution has a firm theoretical basis. This intriguing (but natural from a physical point of view) idea was first presented and investigated  in \cite{early1}, and then it was reinvestigated or rediscovered in a number of early studies on this topics \cite{early2, early3,early4,early5, early6,early7, early8, early9a, silverman2002dark, rotha2002vortices, early9}.

A systematic approach for the in depth investigation of the properties of the Bose-Einstein Condensate dark matter halos was introduced in \cite{BoHa07}, and it is based on the hydrodynamic formulation of the non-relativistic Gross-Pitaevskii (GP) equation, describing the properties of the static dark matter halos in the presence of a confining gravitational potential.  An important advantage of the approach is that the study of the condensate dark matter is significantly simplified due to the introduction of the Madelung (hydrodynamic) representation of the wave function in the Gross-Pitaevskii equation. The hydrodynamic representation allows  to formulate the GP equation in a mathematical form that is similar to classical fluid mechanics. This approach significantly simplifies the application of the model to a wide range of physical situations, thus extending the range of its applications, and also allows a clear and transparent physical interpretation. More exactly, in the hydrodynamic representation, the dynamics of condensate dark matter is described by a standard continuity equation, and a hydrodynamic Euler type equation, which contains a quantum pressure, and a supplementary quantum potential, respectively. With the use of the equivalent Euler equation for the condensate dark matter, confined in a finite volume by the gravitational trapping potential,  one can obtain the basic result that the low temperature bosonic dark matter in a condensate phase can be described from a physical point of view as a non-relativistic gas,  with the pressure satisfying, as a function of density, a polytropic type equation of state, with the polytropic index given by $n=1$. The study of the various properties of the Bose - Einstein Condensate dark matter halos, as well as of their impact on the  cosmological evolution and on the astrophysical properties of the galaxies, is presently a very active and interesting field of research \cite{HaM,Har1,inv0, inv1, inv2,inv3,inv4,inv5,inv6,inv7,inv8,inv9,inv10,inv11,inv12,inv13,inv14,inv15,inv16,inv17,inv18,inv19,inv20,inv21,inv22,inv23,inv24,inv25,inv26, inv27,inv28,inv29,inv30,inv31,inv32,inv33,inv34,inv35, inv37,Chavn, Hui,Chav1, Mem3,Zhang, inv36, inv36a, inv38,inv39, inv40,inv41, inv42, inv43}.

The astrophysical and cosmological implications of the so-called Fuzzy Dark Matter, representing a particular form of dark matter, assumed to exist in the form of an extremely light boson, having a mass of the order of $m\sim 10^{-22}$ eV, and a de Broglie wavelength in the range of $\lambda \sim 1$
kpc, were investigated in \cite{Hui}.
 The properties of the galactic rotation curves in the Bose-Einstein Condensate dark matter model, with quadratic self-interaction, were considered in \cite{inv43}, by using 173 galaxies from the recently published Spitzer Photometry \& Accurate Rotation Curves (SPARC) data.  It turns out that the Bose-Einstein Condensate dark matter model gives a good description of the SPARC data.

From the fundamental result that condensate dark matter can be described as a $n=1$ polytropic gas it follows that the radius $R$ of the {\it static and zero temperature} Bose-Einstein Condensate dark matter
halo can be obtained from the relation by $R=\pi \sqrt{\hbar ^{2}a/Gm^{3}}$, where $a$ is the scattering length \cite{BoHa07}. Hence the two basic physical parameters of the condensate, the mass of the dark matter particle, and its scattering length, fully determine the macroscopic astrophysical properties of the static condensate. The total mass $M$  of the static galactic halo is obtained as $M=4\pi
^2\left(\hbar ^2a/Gm^3\right)^{3/2}\rho _c=4R^3\rho _{c}/\pi $, where by $%
\rho _{c}$ we have denoted the central density of the halo.  The mass of the dark matter particle obeys an interesting mass-galactic radius relation, given by \cite{BoHa07}
\begin{eqnarray}  \label{mass}
m &=&\left( \frac{\pi ^{2}\hbar ^{2}a}{GR^{2}}\right) ^{1/3}\approx
6.73\times 10^{-2}\times \left[ a\left( \mathrm{fm}\right) \right] ^{1/3}\times \nonumber\\
&&\left[ R\;\mathrm{(kpc)}\right] ^{-2/3}\;\mathrm{eV}.
\end{eqnarray}

 By assuming an interparticle  scattering length of the order of $a\approx 10^{-3}$ fm, the above equation points towards a mass of the dark matter particle of the order of a few meV. Very light particle masses of the order of $m\sim 10^{-22}$ eV would require scattering lengths of the order of $a\sim 10^{-60}$ fm.

 An important result in the Bose-Einstein Condensate dark matter approach to the structure of the galactic halos is that their density profiles  indicate the presence of an extended core, whose existence is a result of  the strong interaction
between dark matter particles \cite{Har1}. This feature of the Bose-Einstein Condensate dark matter models naturally solve the famous core-cusp problem, which plagues the standard dark matter models based on the Navarro-Frenk-White density profiles.

Up to now most (if not all) investigations of the Bose-Einstein Condensate dark matter models have been performed under the assumption that the dark bosonic case is {\it ideal}. However, in a realistic physical environment this assumption may not be correct. One of the important physical aspects that may play an important role in the structure of the condensate is the presence of the {\it disorder}, which may be induced by the presence of a {\it random potential}. The study of the effects of the disorder generated by a random potential on the Bose-Einstein Condensates was initiated in \cite{Nat5}, where it was shown that at absolute zero, the random potentials can partially deplete the Bose condensate, and they generate an amount of normal fluid equal to 4/3 of the condensate depletion. The effects of the random potentials on the condensate was investigate in \cite{Nat6,Nat00,Nat0,Nat01,Nat10,Nat20,Nat3,Nat4}. Experimentally, Bose-Einstein Condensates were studied in laboratory in random environments  in the superfluid phase of $^4$He in Vycor glass or aerogels, in $^3$He in aerogels, and in ultracold alkali atoms in disordered traps (see \cite{Nat3} and references therein).

Since the classic studies of Chandrasekhar \cite{C1,C2}, the role of the stochastic processes and methods have been proven to play a fundamental
role in our understanding of many physical and astrophysical phenomena \cite{C3}, which are naturally associated to random behaviors.
There are many astrophysical processes that are random in nature, and can generate a high level of disorder in the cosmic environment. For example, close encounters between stars belonging  to a nuclear cluster, and a massive black hole, can  lead to the disruption of the massive stars due to the extremely strong gravitational tidal forces of the black holes. The gas created in this way can either be dispersed in the space, or accreted by the black hole \cite{C4,C5}. These astrophysical processes related to the stellar collisions or encounters with black holes can generate large amounts of matter that provide a mechanisms for the mass increase of black holes in dense stellar clusters, and for randomly enriching with baryonic matter the dark matter halo. Another source of material is represented by stars moving on highly elongated orbits around black holes, which are prone to tidal disruptions.  Such  events can produce cosmic material in the form of baryons, and their recurrence is essentially random \cite{C4,C5}. Black holes having intermediate masses have a high probability of capturing stars, thus contributing to the tidal energy dissipation, which happens within a time interval of a few core relaxation times \cite{S1,S2}. There are several other sources for disorder in Bose-Einstein Condensates. From an astrophysical point of view an important source of disorder may be represented by the presence of impurities in the condensate at the moment of condensation, and at later times (for example, hydrogen atoms), locally interacting with the condensate, and generating random fluctuations in the condensate distribution \cite{New1}. Even if the initial condensation of dark matter may have taken in the very early Universe, it is still realistic to assume that the cosmic environment did not consist of "pure" dark matter only, and that a many other particle type that did not condense were present when the condensation of the bosonic dark matter occurred. Disorder does appear in particle mixtures \cite{New2}, as well as in condensates located in inhomogeneous magnetic fields.   Spatial heterogeneities as well as a complex topology of the condensing system can also act as sources of disorder \cite{New2}.

It is goal of the present paper to extend the standard Bose-Einstein Condensate dark matter model \cite{BoHa07} by including in the theoretical framework two effects of astrophysical relevance. These are the effects of the baryonic matter on the halo structure, and of the presence of a random potential inside the condensate distribution. A qualitative estimate of the disorder energy shows that it may be of the same order of magnitude as the gravitational or interaction energy, and thus it may play a significant role in the galactic structure. By using the hydrodynamic representation of the Gross-Pitaevskii equation, one can obtain a second order linear differential equation for the condensate density, which also contains the baryonic matter density, and the effective density associated to the disorder potential. By adopting a simple exponential distribution for the baryonic matter density, the density distribution of the condensate dark matter can be obtained exactly by using techniques based on the Laplace transform. Then all the relevant galactic astrophysical properties (mass distribution and tangential velocity) are represented in an exact analytical form. The theoretical model is tested by comparing its predictions with the observational properties  of a set of 39 galaxies from the Spitzer Photomery \& Accurate Rotation Curves (SPARC) database. Estimates of the relevant astrophysical parameters of the dark matter dominated galaxies, including the baryonic matter properties, and the parameters of the random potential, are obtained by fitting the galactic rotation curves. It turns out that the BEC model in the presence of baryonic matter and a random confining potential gives a good description of the observational SPARC data. The presence of the condensate dark matter in the presence of random potentials could also provide a solution to the core/cusp problem.

The present paper is organized as follows. In Section~\ref{sect1} we introduce the Bose-Einstein Condensate dark matter model in the presence of disorder, which is described via a disorder potential, and a corresponding disorder density. With the use of the hydrodynamic representation of the Gross-Pitaevskii equation we obtain the basic equation describing the density of the condensate dark matter halo, and we discuss the effects of the disorder in a qualitative way. The physical properties of the condensate dark matter halos in the presence of a random disorder potential are investigated in Section~\ref{sect2}, where the density, mass and tangential velocity profiles are obtained. We compare the theoretical predictions with the observational data in Section~\ref{sect3}. In Section~\ref{sect4} we discuss and conclude our results.

\section{Bose-Einstein Condensates in random confining potentials}\label{sect1}

In the study of the Bose-Einstein Condensate dark matter models the main theoretical tool is the Hamilton operator of the interacting system of bosons, which is generally can be constructed as \cite{re1, re2,re3,re4, Nat1}
\bea
\hat{H}&=&\int d^3\left(\vec{r}\right)\hat{\Psi}^{+}\left(\vec{r}\right)\Bigg[-\frac{\hbar ^2}{2m}\Delta +U\left(\vec{r}\right)+U_{rot}\left(\vec{r}\right)+\nonumber\\
&&\frac{1}{2}\int d^3\left(\vec{r}\;'\right)U_{int}\left(\vec{r}-\vec{r}\;'\right)\hat{\Psi}^{+}\left(\vec{r}\;'\right)\hat{\Psi}\left(\vec{r}\;'\right)\Bigg]\hat{\Psi}\left(\vec{r}\right),\nonumber\\
\eea
where  by $\hat{\Psi}\left(\vec{r}\right)$ and  $\hat{\Psi}^{+}\left(\vec{r}\;'\right)$  we have denoted the annihilation and creation operators at the position $\vec{r}$, respectively. Furthermore, $m$ is the mass of the particle in the dark matter condensate, while $U$ and $U_{rot}$  denote the external potential, and the effective centrifugal potential, respectively. The  interparticle interaction in the dark halo is described by the potential  $U_{int}$. For the study of the rotational properties of the dark matte halo in the condensed phase we adopt the comoving frame, which is the frame that rotates at the same speed as the galactic dark matter halo.

In the standard theory of the  Bose-Einstein Condensations one usually introduces the basic assumption according o which the interparticle interaction is a short range one. This assumption allows to write the interaction potential as the product between a constant $g$, which depends on the scattering length of the particles, and a Dirac delta function \cite{re1,re2,re3,re4}, so that
\begin{equation}
\begin{aligned}
U_{int}\left(\vec{r}\;'-\vec{r}\right)=&u_0\delta\left(\vec{r}\;'-\vec{r}\right),
\end{aligned}
\end{equation}
where we have denoted
\be
u_0=\frac{4\pi a\hbar^2}m,
\ee
with $a$ representing the scattering length. The scattering length is related to the scattering cross section $\sigma$ by the relation $\sigma =\left(4\pi/{\tilde k}^2\right)\delta _0^2=4\pi a^2$, where ${\tilde k}$ is the wave vector of the scattered wave, and $\delta _0=-{\tilde k}a$ \cite{re3}.

Next we introduce the mean field description, which consists in decomposing the field operator in the Heisenberg picture according to  $\hat{\Psi}(\boldsymbol{r},t)=\psi(\boldsymbol{r},t)+\hat{\Psi}'(\boldsymbol{r},t)$, where $\psi(\boldsymbol{r},t)=\langle\hat{\Psi}(\boldsymbol{r},t)\rangle$ is  called the condensate wave function, and
$\hat{\Psi}'(\boldsymbol{r},t)$ represents the fluctuations in the system. Then, after integrating the Heisenberg equation of motion, we obtain the Gross-Pitaevskii equation, describing the physical properties of a Bose-Einstein Condensate in the adopted approximation, as
\bea\label{gp}
  i\hbar\frac{\partial}{\partial t}\psi(\vec{r},t)&=&\Big[-\frac{\hbar^2}{2m}\nabla^2+U\left(\vec{r}\right)+U_{rot}\left(\vec{r})\right)
  +\nonumber\\
 &&+ u_0|\psi(\vec{r},t)|^2)\Big]\psi\left(\vec{r},t\right).
\eea

The number density $n$ of the dark matter particles  in the condensate is obtained as $n\left(\vec{r},t\right)=|\psi\left(\vec{r},t\right)|^2$. Moreover, the wave function satisfies the normalisation condition is $N=\int {n(\boldsymbol{r},t)d^3\boldsymbol{r}}$.  The mass density of the dark matter condensate  is denoted by $\rho =m n\left(\vec{r},t\right)$.

As for the external potential $U$, we construct it as the sum of the trapping gravitational potential $\phi$, created by the  mass distribution of the dark matter of the halo, of the gravitational effect of the baryonic matter $U_b$, and a {\it random Gaussian potential} $U_{dis}$, which {\it corresponds to the degree of disorder in the system}, and is produced by the random environment, so that
\be
U\left(\vec{r}\right)=\phi \left(\vec{r}\right)+U_b \left(\vec{r}\right)+U_{dis}\left(\vec{r}\right).
\ee

 The external potential $U $ satisfies the Poisson equation,
\be
\Delta U \left(\vec{r}\right)=4\pi Gm\left[\rho \left(\vec{r}\right)+\rho_b\left(\vec{r}\right)+\rho _{dis}\left(\vec{r}\right)\right],
\ee
where $G$ denotes the gravitational constant, $\rho _b$ is the baryonic matter density, and $\rho _{dis}$ denotes the random fluctuations of the matter density in the system.  We characterize  the random potential $U_{dis}$ and the random density fluctuations $\rho _{dis}$ by their average values \cite{Nat1},
\be\label{dis1}
\Big<U_{dis}\left(\vec{r}\right)\Big>=0,\Big<U_{dis}\left(\vec{r}\right)U_{dis}\left(\vec{r}\;'\right)\Big>=\kappa ^2\delta \left(\vec{r}-\vec{r}\;'\right),
\ee
where $\kappa ^2$ is a constant, having the units of g$^2$m$^7$/s$^4$, and
\be
\Big<\rho_{dis}\left(\vec{r}\right)\Big>=0,\Big<\rho_{dis}\left(\vec{r}\right)\rho_{dis}\left(\vec{r}\;'\right)\Big>=\kappa_1 ^2\delta \left(\vec{r}-\vec{r}\;'\right),
\ee
respectively. Hence, we have adopted for the random potential a zero correlation length, as well as a $\delta$-type correlator.  This kind of potential is called {\it uncorrelated random potential} \cite{Nat2}.

\subsection{The hydrodynamic representation}

The hydrodynamic representation of the Gross-Pitaevskii equation is obtained by representing the wave function in the form
\be
\psi\left(\vec{r},t\right)=\sqrt{n\left(\vec{r},t\right)}e^{iS\left(\vec{r},t\right)/\hbar},
\ee
where $S\left(\vec{r},t\right)/\hbar$ is the phase of the wave function. Then, after some simple calculations the Gross-Pitaevskii Eq.~(\ref{gp}) can be reformulated in an equivalent mathematical form as a continuity and hydrodynamic type Euler equation, given by
\be
\frac{\partial n}{\partial t}+\nabla \cdot \left(n\vec{v}\right)=0,
\ee
and
\bea\label{eu1}
m\frac{d\vec{v}}{dt}&=&m\Bigg[\frac{\partial \vec{v}}{\partial t}+\left(\vec{v}\cdot \nabla\right)\vec{v}\Bigg]=-\nabla \Bigg[U\left(\vec{r}\right)+U_{rot}\left(\vec{r}\right)+\nonumber\\
&&u_0n-\frac{\hbar ^2}{2m\sqrt{n}}\Delta \sqrt{n}\Bigg],
\eea
respectively, where the velocity of the condensate is defined according to $\vec{v}=\nabla S\left(\vec{r},t\right)/m$. For a static dark matter distribution,  all terms containing time derivatives in the above equations vanish identically, and we can also neglect the macroscopic motions (the velocity) of the halo. Moreover, we also neglect the last term, describing  the so-called quantum potential, in Eq.~(\ref{eu1}). This approximation is  called Thomas-Fermi approximation. Therefore Eq.~(\ref{eu1}) takes the form
\be\label{13}
 \nabla \Bigg[U\left(\vec{r}\right)+U_{rot}\left(\vec{r}\right)+
u_0n\Bigg]=0.
\ee

 The rotational potential is given by the simple relation
\be
U_{rot}=-\omega^2r^2,
\end{equation}
where $\omega$ is the angular velocity of the dark matter halo. Hence, after applying again the $\nabla$ operator, from Eq.~(\ref{13}) we obtain the basic equation describing the equilibrium properties of the rotating Bose-Einstein Condensate dark matter halos in the presence of a gravitational field and of a random potential,
\be\label{dens}
\Delta \rho \left(\vec{r}\right)+k^2\left[\rho \left(\vec{r}\right)+\rho _b\left(\vec{r}\right)+\rho _{dis}\left(\vec{r}\right)-\frac{\omega ^2}{2\pi G}\right]=0,
\ee
where we have denoted
\be
k^2=\frac{4\pi Gm^2}{u_0}=\frac{Gm^3}{a\hbar ^2}.
\ee
From a mathematical point of view Eq.~(\ref{dens}) represents a second order partial differential Helmholtz type equation.

\subsection{Radii and masses of dark matter Bose-Einstein Condensates in random environments}

In the present Subsection we will investigate, in a simple qualitative way, the effects of the random external potential on the physical properties of the dark matter Bose-Einstein Condensates. The random
potential is characterized by a single parameter $\kappa$. With its help and together with the universal constant $\hbar$ and the mass of the dark matter particle we can construct the length scale  \cite{Nat3, Nat4}
\be
L_{dis}=\frac{16\pi \hbar ^4}{27m^2\kappa ^2},
\ee
which is called the Larkin length, while the corresponding energy scale $\varepsilon$ is given by  \cite{Nat3, Nat4}
\be
\varepsilon =\frac{\hbar^2}{mL_{dis}^2}=\frac{m^3\kappa ^4}{\hbar ^6}.
\ee

In the following we consider a system of $N$ bosons, all in the same quantum state, localized in a volume $V$, with spatial extent $R$, having a total  mass $M$, and confined by the gravitational potential, and the superposed random fluctuations.   The total energy $E$ of the system can be written as
\be
E=E_{kin}+E_{int}+E_{grav}+E_{dis},
\ee
where $E_{kin}$, $E_{int}$, $E_{grav}$ and $E_{dis}$ are the kinetic energy, interaction energy, gravitational energy, and random disorder energy, respectively. The kinetic energy per particle is $\hbar ^2/2mR^2$ \cite{re3}, and hence the total kinetic energy of the system is obtained as $E_{kin}=N\hbar ^2/2mR^2$. The interaction energy is given by $E_{int}=(1/2) \left(N^2/V\right)mu_0$ \cite{re3}, while for the gravitational potential energy we assume the simple expression $E_{grav}=-GM^2/R$. For the disorder energy per particle we adopt the expression \cite{Nat3,Nat4}
\be
\left|E_{dis}\right|=\frac{2\hbar ^2}{3m}\frac{1}{\left(L_{dis}R^3\right)^{1/2}}.
\ee

The above relation can be obtained as follows. According to Eq.~(\ref{dis1}) the average of the disorder potential vanishes.
The variance of the disorder energy in a region of linear extension $R$ follows from the right hand side of Eq.~(\ref{dis1}  after integrating over  both  $\vec{r}$ and $\vec{r}\;'$ over the volume $V=R^3$. The obtained relation represents the typical energy associated with the random potential. Hence we obtain
\bea
E_{dis}^2&\sim & \kappa ^2 \int_{R^3}d^3\vec{r}\int_{{\it R}^3}\Psi^{+}\left(\vec{r}\right)\Psi \left(\vec{r}\right)\delta \left(\vec{r}-\vec{r}\;'\right)\times \nonumber\\
&&\Psi^{+}\left(\vec{r}\;'\right)\Psi \left(\vec{r}\;'\right)d^3\vec{r}^{\;\prime}\sim \kappa ^2\int_{R^3}{d^3\vec{r}\frac{1}{R^6}e^{-\vec{r}^2/R^2}}\sim \nonumber\\
&&\kappa ^2R^{-3}.
\eea

Therefore the total energy of the system can be written as
\be
E=N\frac{\hbar ^2}{2mR^2}+\frac{3}{2}N^2\frac{\hbar ^2a}{mR^3}-\frac{GM^2}{R}\pm \frac{2\hbar ^2}{3m}\frac{N}{\left(L_{dis}R^3\right)^{1/2}}.
\ee

If the condition
\be
\frac{3Na}{R}>>1,
\ee
is satisfied, then the interaction energy is much larger than the kinetic energy, $E_{int}>>E_{kin}$. To prove this result we assume that the mass of the condensate dark matter halo is $M=10^{10}M_{\odot}=2\times 10^{43}$ g, and we take (quite arbitrarily) for the mass of the dark matter particle the value $m=3.19\times 10^{-37}$ g. Then  the number of dark matter particles in the galaxy is of the order $N=M/m=6.26\times 10^{79}$. For a condensate with radius  $R=10$ kpc = $3\times 10^{22}$ cm, the quantity $3(N/R)a=6.26\times 10^{57}a$, which is obviously much greater than one even for unrealistically small values of the scattering length $a$ of the order of let's say $a=10^{-50}$ cm.

Hence the contribution of the kinetic energy of the dark matter particles in the galactic halo can be neglected with respect to the interaction energy of the bosons. This is called the Thomas-Fermi approximation, and in the case of a large particle number, it gives a very good description of the condensate. In the limit of large particle numbers the Thomas-Fermi approximation acquires such a high level of  precision that the corresponding description of the dark matter halos can be considered exact. However, at length scales of the order of $R\approx \sqrt{m/4\pi \rho a}$, where $\rho $ is the mean dark matter density, the Thomas-Fermi approximation breaks down. For example, for $\rho =10^{-24}\;{\rm g/cm^3}$, a density of the same order of magnitude as the galactic densities, we find $R\approx 377.65$ cm, a length scale value that is insignificant from the point of view of the galactic astrophysics.

For the considered numerical values of the parameters the gravitational energy of the dark matter halo has the value
\be
\left|E_{grav}\right|=8.89\times 10^{56}\;{\rm  ergs},
\ee
 while
 \be
 E_{int}=1.35\times 10^{56} \;{\rm ergs}.
 \ee

 This shows that the bosonic interaction energy and the gravitational energy are of the same order of magnitude. However, since $E_{grav}>E_{int}$, it follows that for the condensate dark matter halo the gravitational energy is indeed the trapping energy. As for the disorder energy, by taking into account the explicit form of the Larkin length, it takes the form
\bea
\left|E_{dis}\right|&=&\sqrt{\frac{3}{\pi}} \frac{N\kappa}{2R^{3/2}}=5.89\times 10^{45}\times \nonumber\\
&& \frac{N}{10^{79}}\times \left(\frac{R}{10\;{\rm kpc }}\right)^{-3/2} \kappa \;{\rm ergs}.
\eea

The disorder energy is linearly proportional to $\kappa $, and in the presence of a strong disorder it can exceed the magnitude of the gravitational or interaction energy. The equilibrium size $R$ of the condensate can be determined from the equilibrium condition $\partial E/\partial R=0$. In the Thomas-Fermi approximation, by neglecting the disorder energy, we have
\be
E=\frac{3}{2}N^2\frac{\hbar ^2a}{mR^3}-\frac{GN^2m^2}{R}, \frac{\partial E}{\partial R}=\frac{G m^2 N^2}{R^2}-\frac{9 a h^2 N^2}{2 m R^4},
\ee
giving for the radius of the condensate the expression \cite{BoHa07}
\be
R\sim \sqrt{\frac{a\hbar ^2}{Gm^3}}.
\ee
In the presence of the disorder energy, the equilibrium condition $\partial E/\partial R=0$ gives for the radius of the dark matter halo the algebraic equation
\be
R^2=\frac{9}{2}\frac{a\hbar ^2}{Gm^3}\pm \sqrt{\frac{27}{\pi}}\frac{\kappa R^{3/2}}{2Gm^2N}.
\ee
If the second term in the above equation dominates over the first term, and the disorder energy is positive, the galactic condensate is confined by the random potential energy of the cosmic environment, and its radius is given by
\be
R\sim \frac{27}{4\pi}\frac{\kappa ^2}{G^2m^4N^2}.
\ee
The radius of the galactic halo depends on the particle number, as well as of the characteristics of the random potential, described by $\kappa ^2$. In the first order approximation we obtain for the radius of the condensate the expression
\be
R^2\approx \frac{9}{2}\frac{a\hbar ^2}{Gm^3}\left[1\pm \frac{27}{2^{3/4}\sqrt{\pi}}\left(\frac{G}{a\hbar ^2m^5}\right)^{1/4}\frac{\kappa }{N}\right].
\ee

\section{Weekly disordered dark matter in spherical symmetry}\label{sect2}

In the present Section we will investigate in detail the astrophysical properties of condensate dark matter halos in the presence of uncorrelated disorder. In particular, we will obtain the density and the mass profiles of the halo, and the tangential velocity of massive particle moving around the galactic center.

\subsection{The density profile of the condensate dark matter halo}

In spherical symmetry, by neglecting any angular dependence of the physical parameters,  Eq.~(\ref{dens}) becomes
\begin{equation}
\frac{1}{r^{2}}\frac{d}{dr}\left( r^{2}\frac{d\rho }{dr}\right) +k^2\left( \rho +\rho _{b}+\rho _{dis}-\frac{\omega ^2}{2\pi G}\right) =0,
\end{equation}%
or, equivalently,
\begin{equation}
\frac{d^{2}\rho }{dr^{2}}+\frac{2}{r}\frac{d\rho }{dr}+k^2\left( \rho +\rho _{b}+\rho _{dis}-\frac{\omega ^2}{2\pi G}\right) =0.  \label{2}
\end{equation}

In the following we rescale the variable by means of the transformations
\begin{equation}
r=\frac{\xi }{k},\rho =\rho _{c}\Theta ,
\end{equation}%
where $\rho _{c}$ is the central density of the dark matter halo. Moreover, we
denote
\bea
\Omega ^{2}&=&\frac{\omega ^{2}}{2\pi G\rho _{c}}=0.02386\times \left(\frac{\omega}{10^{-16}\;{\rm s}^{-1}}\right)^2\times \nonumber\\
&&\left(\frac{\rho _c}{10^{-24}\;{\rm g/cm^3}}\right)^{-1}.
\eea

Therefore Eq. (\ref{2}) takes the dimensionless form
\begin{equation}
\xi \frac{d^{2}\Theta }{d\xi ^{2}}+2\frac{d\Theta }{d\xi }+\xi
\Theta =\left[ \Omega ^{2}-\frac{\rho _{b}\left( \xi \right) }{\rho _{c}}-\frac{\rho _{dis}\left( \xi \right) }{\rho _{c}}%
\right] \xi .  \label{3}
\end{equation}

Eq. (\ref{3}) must be integrated with the initial conditions $\Theta \left(
0\right) =1$ and $\Theta ^{\prime }\left( 0\right) =0$. In order to solve
Eq. (\ref{3}) we adopt a technique based on the use of the Laplace
transform, defined as
\begin{equation}
\mathcal{L}_{\xi}\left[ \Theta \left( \xi \right) \right] \left( s\right)
=F(s)=\int_{0}^{\infty }\Theta \left( \xi \right) e^{-s\xi }d\xi
,s\in \mathbb{R} .
\end{equation}

Then we immediately obtain
\begin{eqnarray}
&&\mathcal{L}_{\xi}\left[ \xi \frac{d^{2}\Theta \left( \xi \right) }{d\xi ^{2}}%
\right] \left( s\right) =-\frac{d}{ds}\int_{0}^{\infty }\frac{d^{2}\Theta
\left( \xi \right) }{d\xi ^{2}}e^{-s\xi }d\xi =  \label{t1} \nonumber\\
&&-\frac{d}{ds}\left( s^{2}F(s)-s\Theta \left( 0\right) -\Theta ^{\prime
}(0)\right) =  \notag \\
&&-s^{2}F^{\prime }\left( s\right) -2sF(s)+\Theta \left( 0\right) ,
\end{eqnarray}%
\begin{equation}
\mathcal{L}_{\xi}\left[ \xi \Theta \left( \xi \right) \right] \left( s\right) =-\frac{d%
}{ds}\int_{0}^{\infty }\Theta \left( \xi \right) e^{-s\xi }d\xi
=-F^{\prime }(s).  \label{t2}
\end{equation}

Since $\mathcal{L}_{\xi}\left[ d\Theta /d\xi \right] \left( s\right) =sF(s)-\Theta \left(
0\right) $, after taking the Laplace transform of Eq. (\ref{3}), we obtain
the first order linear differential equation for $F(s)$ given by
\begin{equation}
F^{\prime }(s)=-\frac{1}{1+s^{2}}-\frac{\Omega ^{2}}{s^{2}\left(
1+s^{2}\right) }+G(s),  \label{4}
\end{equation}%
where we have denoted
\begin{equation}
G(s)=\frac{1}{1+s^2}\left\{\mathcal{L}\left[ \xi \frac{\rho _{b}\left( \xi \right) }{\rho _{c}}\right]+\mathcal{L}\left[ \xi \frac{\rho _{dis}\left( \xi \right) }{\rho _{c}}\right]
\left( s\right)\right\} .
\end{equation}

The general solution of Eq. (\ref{4}) is given by
\begin{equation}
F(s)=\frac{\Omega ^{2}}{s}-\left( 1-\Omega ^{2}\right) \arctan s+\int
G(s)ds.
\end{equation}

The inverse Laplace transform of the function $\arctan s$ can be obtained as
follows. By taking into account that $d(\arctan s)/ds=1/\left(
1+s^{2}\right) $, we obtain
\begin{eqnarray}
\arctan s =\frac{\pi}{2}-\int_s^{\infty}{\frac{ds}{1+s^2}},
\end{eqnarray}
which, by taking into account that $\mathcal{L}_{\xi}^{-1}\left(\pi/2\right)=(\pi /2)\xi$, and $\mathcal{L}_{\xi}^{-1}\left[\int_s^{\infty}{\frac{ds}{1+s^2}}\right]=\sin \xi/\xi$,  gives immediately $\mathcal{L}^{-1}_{\xi}\left[ \arctan s\right] \left( \xi \right)
=(\pi /2)\delta (\xi)-\sin \xi /\xi $, where $\delta (\xi)$ is the Dirac function. Therefore the general solution of Eq. (\ref{3}) can
be obtained as follows
\begin{eqnarray}\label{45n}
\Theta \left( \xi \right) &=&\left(1-\Omega ^2\right)\frac{\sin
\xi }{\xi }+\Omega ^{2}- \left(1-\Omega ^2\right)\frac{\pi}{2}\delta (\xi)+ \nonumber \\
&&\mathcal{L}^{-1}_{\xi}\Bigg\{ \int \frac{1}{1+s^2}\mathcal{L}\left[ \xi \frac{\rho _{b}\left( \xi \right) }{%
\rho _{c}}\right] \left( s\right) ds+\nonumber\\
&&\int \frac{1}{1+s^2}\mathcal{L}_{\xi}\left[ \xi \frac{\rho _{dis}\left( \xi \right) }{%
\rho _{c}}\right] \left( s\right) ds\Bigg\} (\xi)=\nonumber\\
&&\left(1-\Omega ^2\right)\frac{\sin
\xi }{\xi }+\Omega ^{2}+\Theta _{0}\left(\xi\right),
\end{eqnarray}
where
\bea
\Theta _{0}\left(\xi\right)&=&\mathcal{L}^{-1}_{\xi}\Bigg\{ \int \frac{1}{1+s^2}\mathcal{L}_{\xi}\left[ \xi \frac{\rho _{b}\left( \xi \right) }{%
\rho _{c}}\right] \left( s\right) ds+\nonumber\\
&&\int \frac{1}{1+s^2}\mathcal{L}\left[ \xi \frac{\rho _{dis}\left( \xi \right) }{%
\rho _{c}}\right] \left( s\right) ds\Bigg\} (\xi)=\nonumber\\
&&\Theta _0^{(bar)}(\xi)+\Theta _0^{(dis)}(\xi).
\eea

For the case of the nonrotating dark matter halo $\Omega =0$. By neglecting the contributions of the baryonic matter and of the random potential, the static Bose-Einstein condensate dark matter profile can be reobtained immediately as
\be\label{47}
\rho (r)=\rho _c\frac{\sin kr}{kr},
\ee
giving for the radius $R$ of the static condensate the expression \cite{BoHa07}
\bea
R&=&\frac{\pi}{k}=\pi \sqrt{\frac{a\hbar ^2}{Gm^3}}=13.5\times \left(\frac{a}{10^{-17}\;{\rm cm}}\right)^{1/2} \times \nonumber\\
&&\left(\frac{m}{10^{-36}\;{\rm g}}\right)^{-3/2}\;{\rm kpc}  .
\eea

In order to obtain an analytical representation of the function $\Theta _{0}\left(\xi\right)$ we need to specify the baryonic matter density distribution, as well as the specific form of the disorder density. For the sake of simplicity we assume that the baryonic matter profile can be described by a simple exponential function, so that
\be\label{49}
\rho _b(r)=\rho _{bc}e^{- r/r_b},
\ee
where $\rho _{bc}$ is the central baryonic density, and $r_b $ is a constant, which describes the extension of the baryonic component of the galaxy. In the dimensionless variable $\xi$ we have
\be
\rho _b(\xi)=\rho _{bc}e^{-\gamma \xi},
\ee
where $\gamma =R/\pi r_b$. Hence we successively obtain
\be
\mathcal{L}_{\xi}\left[ \xi \frac{\rho _{b}\left( \xi \right) }{%
\rho _{c}}\right] \left( s\right)=\frac{\rho_{bc}}{\rho_c}\frac{1}{\left(s+\gamma \right)^2},
\ee
and
\bea
&&\int \frac{1}{1+s^2}\mathcal{L}_{\xi}\left[ \xi \frac{\rho _{b}\left( \xi \right) }{%
\rho _{c}}\right] \left( s\right) ds=
\frac{\rho_{bc}}{\rho _c}\frac{1}{\left(1+\gamma ^2\right)^2}\times \nonumber\\
&&\Bigg[-\frac{1+\gamma ^2}{s+\gamma}+\left(\gamma ^2-1\right)\arctan s+\ln\frac{(s+\gamma)^{2\gamma}}{\left(1+s^2\right)^{\gamma}}\Bigg],
\eea
respectively, giving for the total baryonic contribution to the condensate dark matter profile the expression
\bea
\hspace{-0.5cm}&&\Theta _0^{(bar)}(\xi)=\frac{\rho _{bc} }{\left(\gamma ^2+1\right)^2  \rho _{c}}  \Bigg\{-\left(\gamma   ^2+1\right)  e^{-\gamma  \xi   }+\nonumber\\
\hspace{-0.5cm}&&\frac{2 \gamma
   \left(\cos \xi   -e^{-\gamma  \xi   }\right)}{\xi   }+\left(\gamma ^2-1\right)   \left[\frac{\pi  \delta   (\xi )}{2}-\frac{\sin \xi
   }{\xi }\right]\Bigg\}.
\eea

The baryonic matter contribution to the dark matter profile satisfies the condition $\lim_{\xi\rightarrow 0}\Theta _0^{(bar)}(\xi)=0$.

To quantify the effect of the random fluctuations on the condensate dark matter profile, we need to specify the functional form of $\rho _{dis}(\xi)$, which we will assume as having a Gaussian distribution, given by
\be
\rho _{dis}(\xi)=Ae^{-\sigma \xi ^2},
\ee
where $A$ and $\sigma $ are constants. Hence we obtain
\be
\mathcal{L}_{\xi}\left[ \xi \frac{\rho _{dis}\left( \xi \right) }{%
\rho _{c}}\right] \left( s\right)=
\frac{A}{2 \sigma \rho_c}-\frac{\sqrt{\pi } A }{4 \sigma ^{3/2}\rho _c}s  e^{\frac{s^2}{4 \sigma }}\text{erfc}\left(\frac{s}{2\sqrt{\sigma }}\right),
\ee
where $\text{erfc}(z)$ denotes the complementary error function, $\text{erfc}(z)=1-\text{erf}(z)$, and
\bea
\int \frac{1}{1+s^2}\mathcal{L}_{\xi}\left[ \xi \frac{\rho _{b}\left( \xi \right) }{%
\rho _{c}}\right] \left( s\right) ds&=&\frac{A}{2 \sigma \rho_c}\arctan s-\nonumber\\
&&\frac{\sqrt{\pi } A }{4 \sigma ^{3/2}\rho _c}I(s),
\eea
respectively, where we have denoted
\be
I(s)=\frac{ s
   e^{\frac{s^2}{4 \sigma }}
   \text{erfc}\left(\frac{s}{2
   \sqrt{\sigma }}\right)}{4
   \left(s^2+1\right) }.
\ee

Therefore for  $\Theta _0^{(dis)}(\xi)$ we obtain
\bea
\Theta _0^{(dis)}(\xi)&=&\frac{A}{2\sigma \rho _c}\frac{\pi}{2}\delta (\xi)-\frac{A}{2\sigma \rho _c}\frac{\sin \xi}{\xi}-\nonumber\\
&&\frac{\sqrt{\pi } A }{4 \sigma ^{3/2}\rho _c}\mathcal{L}^{-1}_{\xi}[I(s)](\xi).
\eea

To obtain an analytical estimate of the role of the dissipation density on the dark matter halo we perform first an asymptotic expansion of $I(s)$, thus obtaining
\be
I(s)\approx \frac{6 \sigma ^{5/2}}{\sqrt{\pi } s^6}-\frac{\left(s^2-1\right) \sigma
   ^{3/2}}{\sqrt{\pi } s^6}+\frac{\left(s^4-s^2+1\right) \sqrt{\sigma }}{2
   \sqrt{\pi } s^6}+...,
\ee
and thus
\bea
\mathcal{L}^{-1}_{\xi}[I(s)](\xi)&\approx &\frac{  \sqrt{\sigma }}{2 \sqrt{\pi }}\xi-\frac{ \sqrt{\sigma } (2 \sigma +1)}{12 \sqrt{\pi
   }}\xi ^3+\nonumber\\
  && \frac{ \sqrt{\sigma } \left(12 \sigma ^2+2 \sigma +1\right)}{240
   \sqrt{\pi }}\xi ^5+....
\eea

By introducing the parameter $\theta =A/2\rho _c$, we obtain for the density of the condensate the expression
\bea
\Theta (\xi)&\approx&   \left(1-\frac{\left(\gamma   ^2-1\right) \rho _ {bc}}{\left(\gamma   ^2+1\right)^2 \rho_{c}}-\theta -\Omega   ^2\right)\frac{\sin \xi }{\xi   }+\Omega ^2-\nonumber\\
&&\frac{\rho _{bc}   e^{-\gamma  \xi }}{(\gamma   ^2 +1)\rho_{c}}+\frac{2 \gamma   \rho_{bc} \left(\cos   \xi -e^{-\gamma  \xi   }\right)}{\left(\gamma   ^2+1\right)^2 \xi  \rho_{c}}+\nonumber\\
&&\frac{1}{2} \pi     \left[\frac{\left(\gamma   ^2-1\right) \rho_{bc}}{\left(\gamma   ^2+1\right)^2\rho_{c}}+\theta -\left(1-\Omega ^2\right)  \right] \delta (\xi )+\nonumber\\
&&\frac{\theta   \xi  \left(\xi ^4 \left(12   \sigma ^2+2 \sigma   +1\right)-20 \xi ^2 (2   \sigma +1)+120\right)}{480   \sigma }.\nonumber\\
\eea
At the center of the halo the function $\Theta (\xi)$ satisfies the condition $\lim_{\xi\rightarrow 0}\Theta (\xi)=1-\theta$. The properties of the dark matter halo are determined by the set of independent parameters $\left(\sigma, \theta, \gamma, \rho_{bc},\rho _c,\Omega\right)$, where $(\sigma, \theta)$ describe the properties of the Gaussian random potential, $\left(\gamma, \rho_{bc}\right)$ describe the effect of the baryonic matter on the dark matter halo, $\rho_c$ is an intrinsic property of the condensate dark matter, while $\Omega $ describes the rotation of the halo.   The variation of the dark matter halo density is represented, for different values of the model parameters, in Fig.~1.

\begin{figure*}[tbp!]
\centering
\includegraphics[width=8cm]{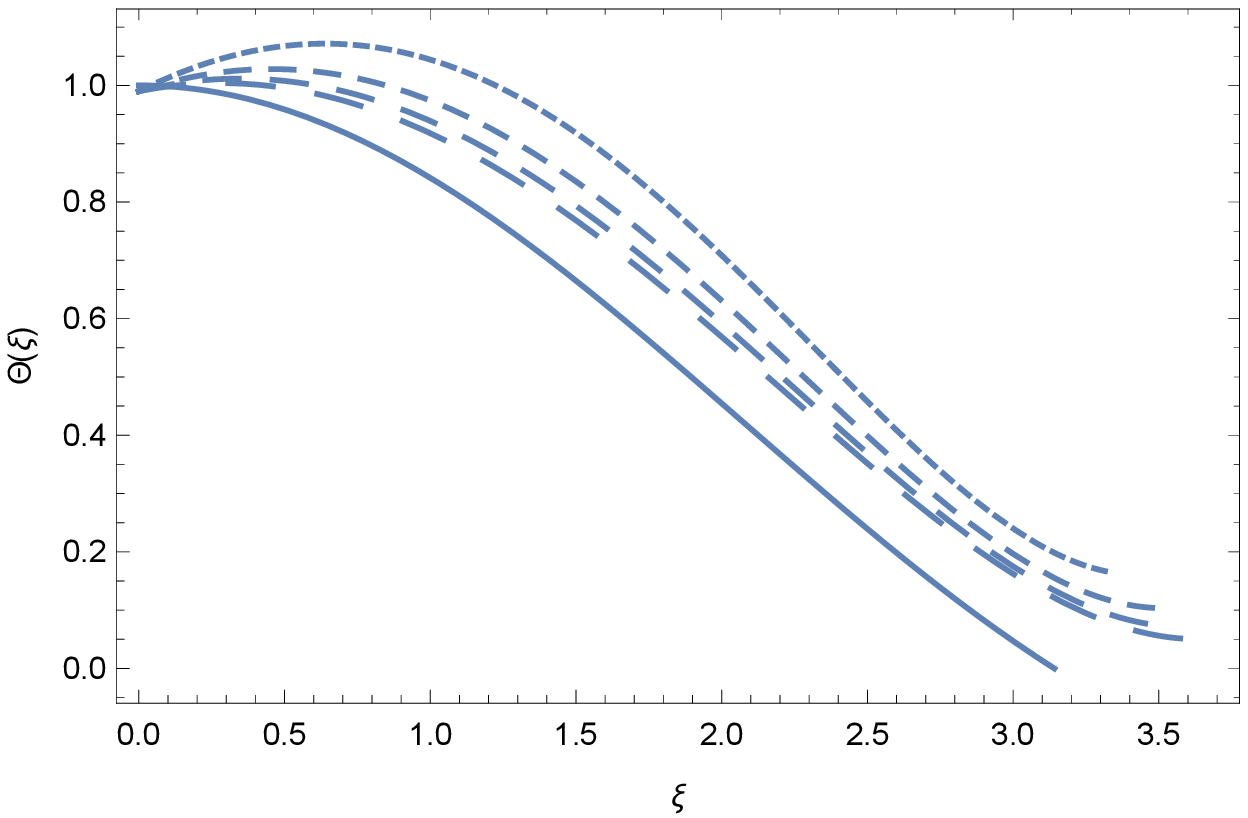}
\includegraphics[width=8cm]{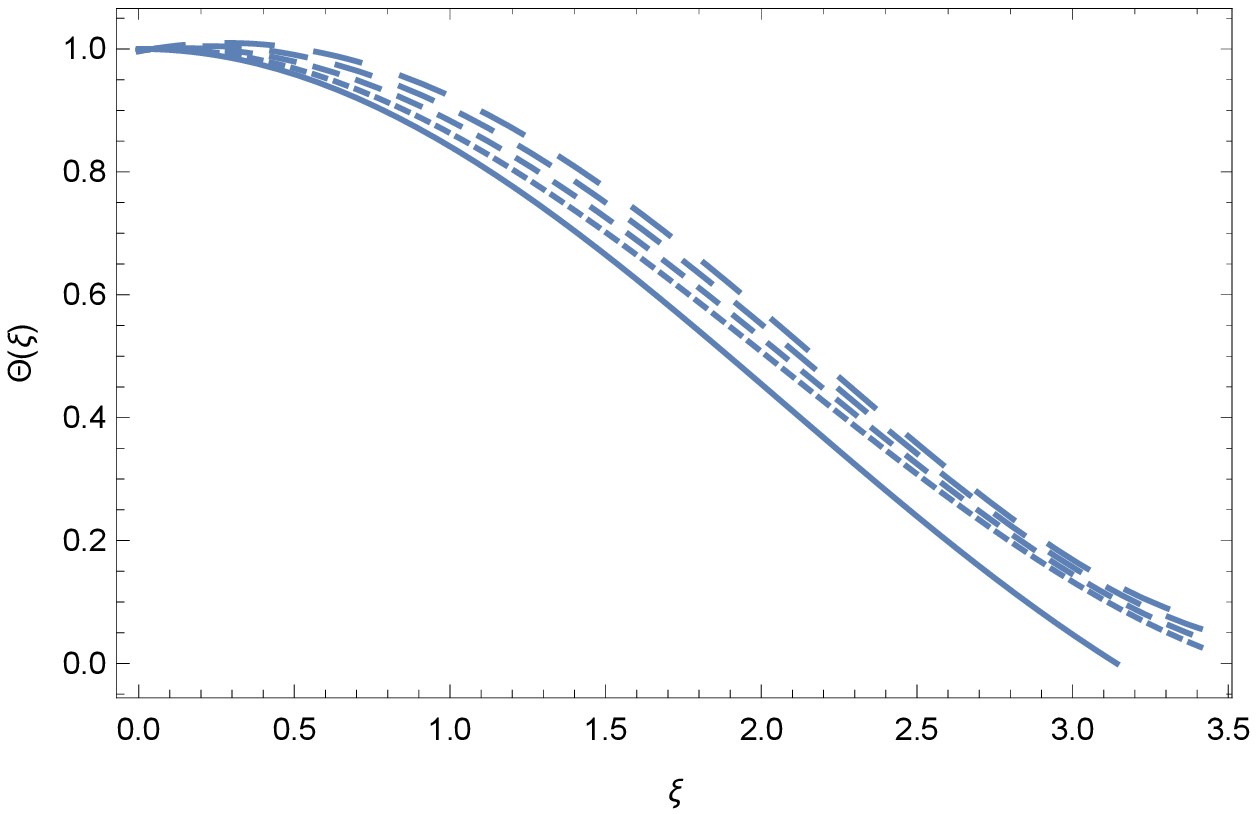}
\caption{Variation of the dimensionless dark matter density $\Theta (\xi)$ as a function of the dimensionless radial distance $\xi$, for different model parameters. In the left Figure, $\theta =0.01$, $\gamma =2$, $\Omega =0.3$, $\rho _{bc}/\rho _c=0.2$, while $\sigma =0.01$ (dotted curve), $\sigma =0.015$ (short dashed curve), $\sigma =0.02$ (dashed curve), and $\sigma=0.025$ (long dashed curve), respectively. In the right Figure, $\sigma =0.01$, $\gamma =2$, $\Omega =0.3$, $\rho _{bc}/\rho _c=0.2$, while $\theta =0.001$ (dotted curve), $\theta =0.002$ (short dashed curve), $\theta =0.003$ (dashed curve), and $\theta=0.004$ (long dashed curve), respectively.  In both Figures the solid curve corresponds to the ideal non-rotating condensate dark matter halo, with $\theta =0$, $\gamma =0$, $\Omega =0$, $\rho _{bc}=0$, and $\sigma =0$.}
\label{fig1}
\end{figure*}

There is a tight correlation between the density Bose-Einstein Condensate dark matter profile and the baryonic and the random density profiles, respectively.   The galactic halo dark matter distribution can be approximated by an expression of the form
\be\label{45}
\Theta \left( \xi \right) \approx \left(1-\Omega ^2\right)\frac{\sin\left(\xi +\xi _{0}\right)}{\xi}= \left(1-\Omega ^2\right)\frac{\sin \xi_{eff}}{\xi}.
\ee
To determine the expression of $\xi _0$ we take into account that $\sin\left(\xi +\xi_{0}\right)=\sin \xi \cos \xi_{0}+\cos \xi \sin \xi_{0}$, and assuming that $\xi _{0}$ is small, we have $\sin\left(\xi +\xi_{0}\right)\approx \sin \xi+\xi_{0}\cos \xi $, giving
\be
\xi_{0}=\frac{\Omega ^2+\Theta _{0}\left(\xi\right)}{\left(1-\Omega ^2\right)\cos \xi}\xi,
\ee
and
\be
\xi_{eff}=\left[1+\frac{\Omega ^2+\Theta _{0}\left(\xi\right)}{\left(1-\Omega ^2\right)\cos \xi}\right]\xi,
\ee
respectively.
Eq.~(\ref{45}) allows us to define the effective radius $\xi_{S}$ of the condensate in the presence of a random potential and of baryonic matter as a solution of the equation $\xi_S+\xi_{eff}\left(\xi_S\right) =\pi$, or equivalently $R_{S}=R\left[1-\left(1\pi\right)\xi_{eff}\left(\pi R_S/R\right)\right]$. Hence one obtains from this equation $R_S$, one can estimate the combined effect of the rotation, baryonic matter, and of the random potential on the dark matter halo.

\subsection{The mass and velocity profiles}

The mass distribution $m(r)$ of the dark matter halo is defined as
\be
m(r)=4\pi \int_0^r{\rho (r)r^2dr},
\ee
where we have introduced the simplifying assumption that the independent effects of the baryonic matter and of the random potential can be approximated through their presence in the dark matter density profile, Eq.~(\ref{45n}). A more detailed investigation would require the construction of explicit baryonic and random density mass profiles, which should then be added as supplementary terms in the total mass of the galaxy. However, in the following we assume that Eq.~(\ref{45n}) contains all the effects related to condensate dark matter, baryonic matter, random potential, and rotation, respectively. In the dimensionless variables $\left(\xi, \Theta (\xi)\right)$ the mass distribution ca then be written as
\bea
m(\xi)&=&\frac{4\rho _cR^3}{\pi^2}\int_0^{\xi}{\Theta (\xi)\xi^2d\xi}=\frac{4\rho _cR^3}{\pi^2}\mu(\xi)=5.47\times 10^9\times\nonumber\\
&& \left(\frac{\rho _c}{10^{-24}\;{\rm g/cm^3}}\right)\left(\frac{R}{10\;{\rm kpc}}\right)^3\times \mu (\xi)M_{\odot},
\eea
where $\mu (\xi)=\int_0^{\xi}{\Theta (\xi)\xi ^2d\xi}$. For $\mu (\xi)$ we obtain, after integrating Eq.~(\ref{45}), the expression
\begin{widetext}
\bea
\mu (\xi)&=&-\frac{2 \rho_{bc}}{\gamma ^3 \rho_{c}}+\sin \xi    \left[\frac{\rho_{bc} \left(-\gamma ^2+2   \gamma  \xi
   +1\right)}{\left(\gamma    ^2+1\right)^2 \rho_{c}}-\theta -\Omega   ^2+1\right]
   +\frac{\cos \xi   \left[\left(\gamma   ^2+1\right)^2 \xi
   \rho _{c}   \left(\theta +\Omega   ^2-1\right)+\left(\gamma   ^2-1\right) \xi  \rho _{bc}+2 \gamma   \rho_{bc}\right]}{\left(\gamma
   ^2+1\right)^2 \rho_{c}}+\nonumber\\
  && \frac{\rho _{bc}   e^{-\gamma  \xi }   \left[\gamma  (\gamma  \xi   +2) \left(\gamma ^2 \xi +2   \gamma +\xi
   \right)+2\right]}{\gamma ^3   \left(\gamma ^2+1\right)^2   \rho_{c}}+\frac{\theta  \xi ^4   \left\{\xi ^4 \left[6 \sigma  (6   \sigma +1)+3\right]-80 \xi ^2 (2   \sigma   +1)+720\right\}}{11520   \sigma }+\frac{\xi ^3   \Omega ^2}{3}.
\eea
\end{widetext}

\begin{figure*}[tbp!]
\centering
\includegraphics[width=8cm]{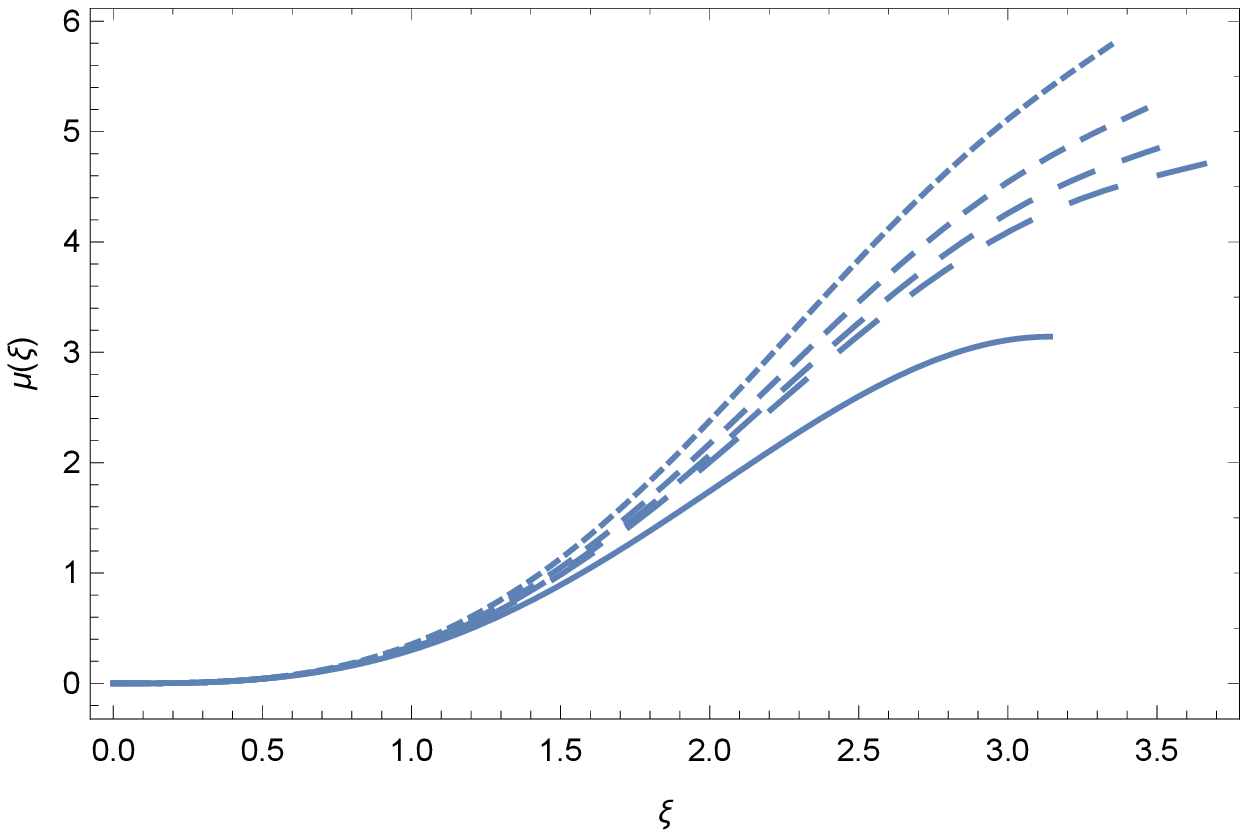}
\includegraphics[width=8cm]{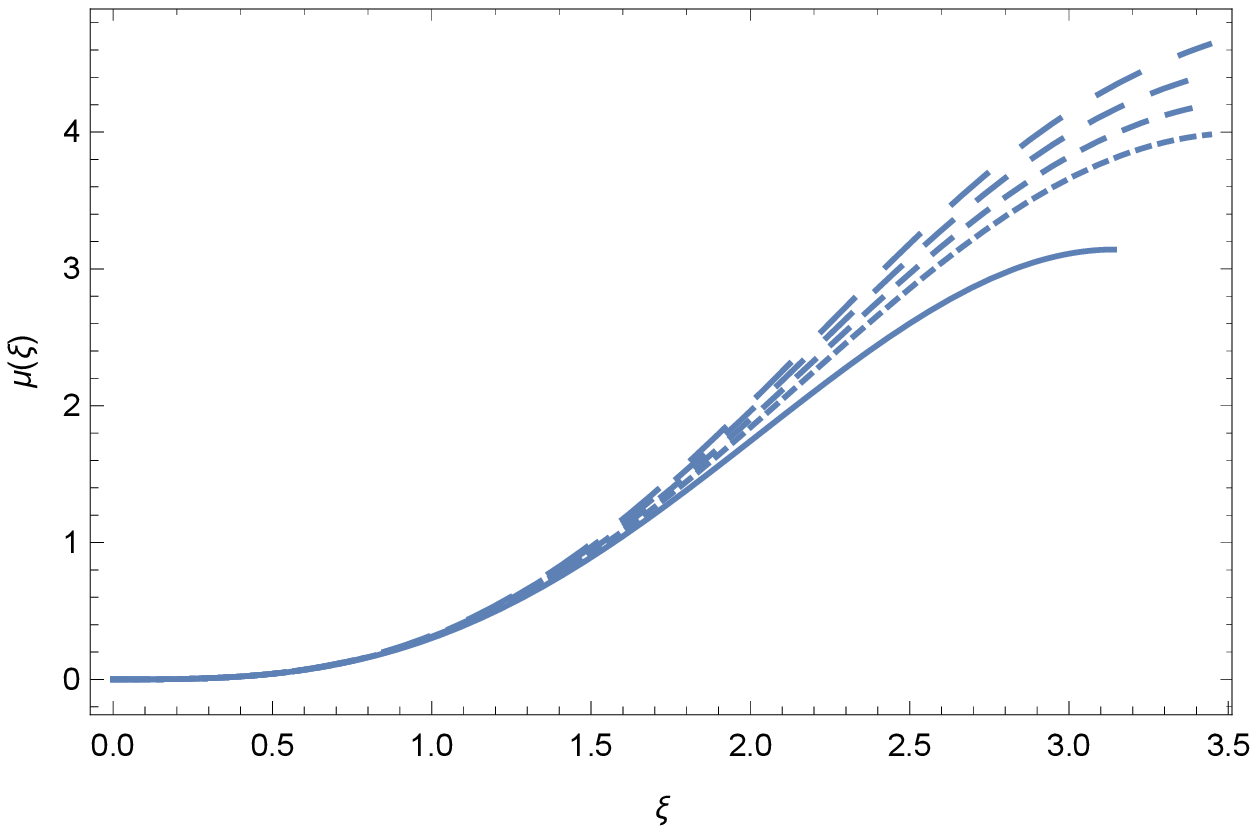}
\caption{Variation of the dimensionless mass $\mu (\xi)$ as a function of the dimensionless radial distance $\xi$, for different model parameters. In the left Figure, $\theta =0.01$, $\gamma =2$, $\Omega =0.3$, $\rho _{bc}/\rho _c=0.2$, while $\sigma =0.01$ (dotted curve), $\sigma =0.015$ (short dashed curve), $\sigma =0.02$ (dashed curve), and $\sigma=0.025$ (long dashed curve), respectively. In the right Figure, $\sigma =0.01$, $\gamma =2$, $\Omega =0.3$, $\rho _{bc}/\rho _c=0.2$, while $\theta =0.001$ (dotted curve), $\theta =0.002$ (short dashed curve), $\theta =0.003$ (dashed curve), and $\theta=0.004$ (long dashed curve), respectively.  In both Figures the solid curve corresponds to the ideal non-rotating condensate dark matter halo, with $\theta =0$, $\gamma =0$, $\Omega =0$, $\rho _{bc}=0$, and $\sigma =0$.}
\label{fig2}
\end{figure*}

The mass distribution has the property $\lim_{\xi\rightarrow 0} m(\xi)=0$. The variation of the function $\mu(\xi)$ is represented, for different values of the parameters of the model, in Fig.~\ref{fig2}.

We define the tangential velocity of a massive test particle in the dark matter halo according to the Keplerian velocity law,
\be
v^2_{DM}(r)=\frac{Gm(r)}{r},
\ee
and
\bea\label{69}
\hspace{-0.5cm}v_{DM}^2(r)&=&\left(\frac{4 G\rho_cR^2}{\pi}\right)\frac{\mu (\pi r/R)}{\pi r/R}=80.56\times \nonumber\\
\hspace{-0.5cm}&&\left(\frac{\rho _c}{10^{-24}\;{\rm g/cm^3}}\right)\times
 \left(\frac{R}{\;{\rm kpc}}\right)^2\times V^2\left(\frac{\pi r}{R}\right)\;{\rm \frac{km}{s}},\nonumber\\
\eea
where we have denoted
\be
V^2(r)=\frac{\mu \left(\pi r/R\right)}{\left(\pi r/R\right)}.
\ee

\begin{figure*}[tbp!]
\centering
\includegraphics[width=8cm]{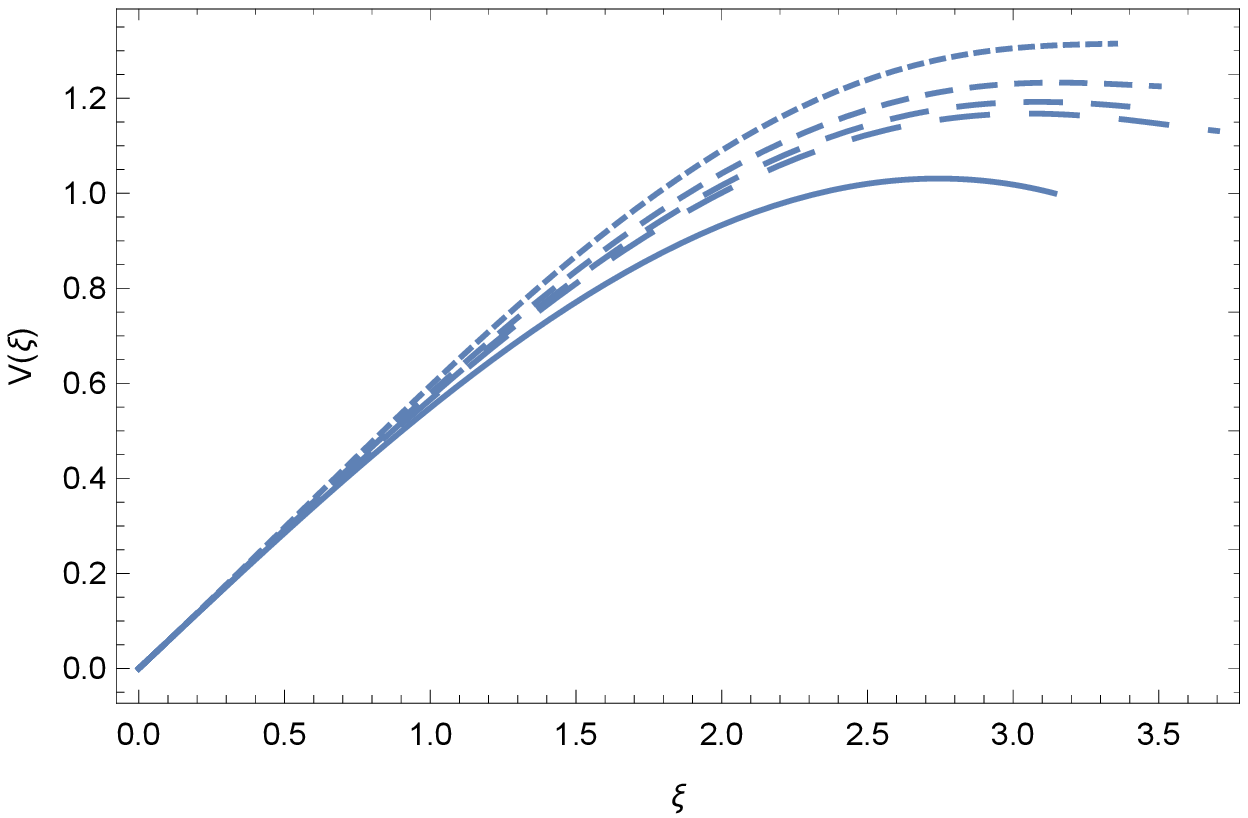}
\includegraphics[width=8cm]{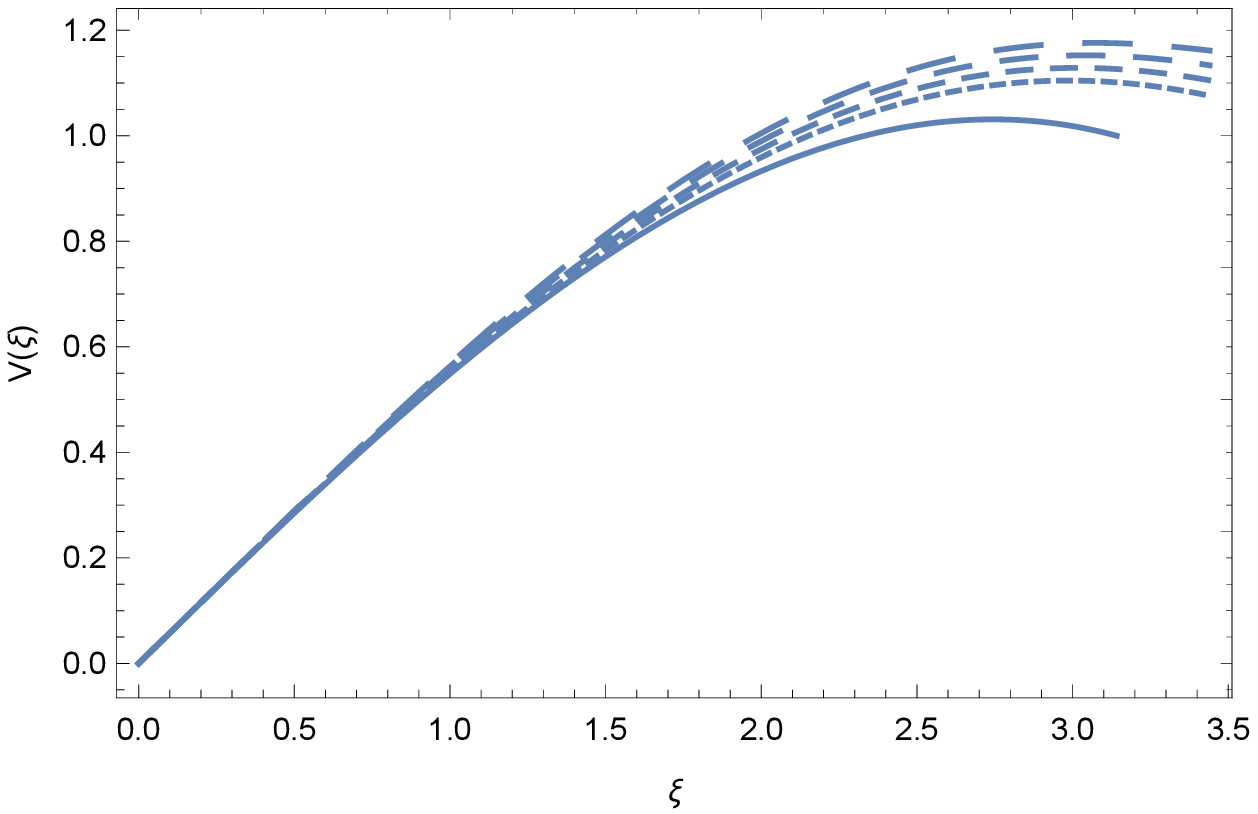}
\caption{Variation of the dimensionless tangential velocity  $V (\xi)$ as a function of the dimensionless radial distance $\xi$, for different model parameters. In the left Figure, $\theta =0.01$, $\gamma =2$, $\Omega =0.3$, $\rho _{bc}/\rho _c=0.2$, while $\sigma =0.01$ (dotted curve), $\sigma =0.015$ (short dashed curve), $\sigma =0.02$ (dashed curve), and $\sigma=0.025$ (long dashed curve), respectively. In the right Figure, $\sigma =0.01$, $\gamma =2$, $\Omega =0.3$, $\rho _{bc}/\rho _c=0.2$, while $\theta =0.001$ (dotted curve), $\theta =0.002$ (short dashed curve), $\theta =0.003$ (dashed curve), and $\theta=0.004$ (long dashed curve), respectively.  In both Figures the solid curve corresponds to the ideal non-rotating condensate dark matter halo, with $\theta =0$, $\gamma =0$, $\Omega =0$, $\rho _{bc}=0$, and $\sigma =0$.}
\label{fig3}
\end{figure*}

The behavior of the function $V(\xi)=\sqrt{\mu (\xi)/\xi}$ is presented, for different values of the model parameters, in Fig.~\ref{fig3}.

\subsubsection{The static BEC condensate}

In the case of a static condensate, with the effects of random fluctuations ignored, the density distribution is given by Eq.~(\ref{47}). The radius of the static halo is defined as the value of the radial coordinate when the density vanishes, $\rho (R)=0$, which gives $k=\pi/R$. Hence for the density of the static halo we obtain the equivalent expression
\be
\rho (r)=\rho _c\frac{\sin \pi r/R}{\pi r/R}.
\ee
For the mass distribution $m(r)=4\pi \int_0^r{r^2\rho (r)dr}$ we obtain
\be\label{72}
m(r)=\frac{4\rho_cR^3}{\pi ^2}\left[\sin \frac{\pi r}{R}-\frac{\pi r}{R}\cos \frac{\pi r}{R}\right].
\ee
The total mass of the condensate is given by $M=m(R)=4\rho _c R^3/\pi$. The tangential velocity of the test particle moving in circular orbits around the galactic center in the condensate dark matter halo, given by $v_{DM}^2=Gm(r)/r$, is obtained as
\be
v_{DM}^2(r)=\frac{4G\rho_c R^2}{\pi }\left[\frac{R}{\pi r}\sin \frac{\pi r}{R}- \cos \frac{\pi r}{R}\right],
\ee
or
\bea\label{73}
\hspace{-0.5cm}v_{DM}^2(r)&=&80.56\times \frac{\rho _c}{10^{-24}\;{\rm g/cm^3}}\times \nonumber\\
\hspace{-0.5cm}&&\left(\frac{R}{{\rm kpc}}\right)^2\times\left[\frac{R}{\pi r}\sin \frac{\pi r}{R}- \cos \frac{\pi r}{R}\right]\;\frac{{\rm km^2}}{{\rm s^2}}.
\eea

\section{Comparison with observational data}\label{sect3}

To compare the velocity of the massive test particles moving on circular orbits in the Bose-Einstein Condensate dark matter halo, as given by Eq.~ (\ref{69}), we fit it with the total observational velocity of the hydrogen clouds that can be obtained from the SPARC database,
and which can be obtained from the relation
\bea\label{v_obs}
&&v_{obs}= \nonumber\\
&&\sqrt{v_{gas}|v_{gas}|\!+\!\Upsilon_d\!\times\! v_{disk}|v_{disk}|\!+\!\Upsilon_b\!\times\! v_{bulge}|v_{bulge}|\!+\!v_{DM}^{2}},\nonumber\\
\eea
where $v_{obs}$ is the total velocity, including the contributions of both baryonic and dark matter, and $v_{gas}$, $v_{disk}$, $v_{bulge}$ and $v_{DM}$ denote the contributions from the velocity of the gas, of the disk, of the bulge, and of the dark matter halo, respectively. Moreover, $\!\Upsilon_d\!$ and $\!\Upsilon_b\!$ denote the  stellar mass-to-light ratios for the disk and the stellar bulge, respectively.

The Spitzer Photomery \& Accurate Rotation Curves (SPARC) database is the largest galactic database obtained up to now, and
that for every galaxy it contains the rotation curves for a large number of galaxies, and it includes
spatially resolved data on the distribution of stars and gas. The range of the rotation velocities is $V_{tg}$ is $20<V_{tg}<300$ km/s, while the luminosities vary between $10^7<L_{[3.6]}<5\times 10^{11}L_{\odot}$. The SPARC database contains information about both large and small mass galaxies. In the SPARC database galaxies with low surface brightness are also well represented.

\subsection{Galaxies selection and fitting algorithm}

\begin{figure}[tbp!]
\centering
\includegraphics[width=8cm]{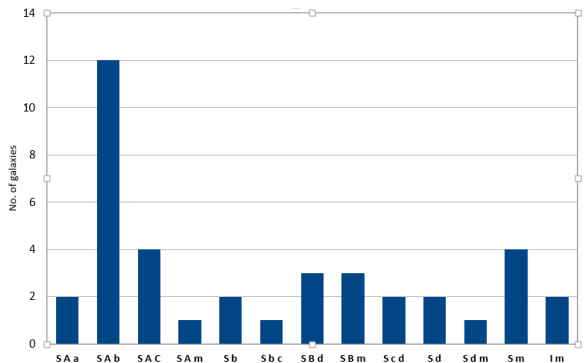}
\caption{Galaxy distribution according to their morphological classification scheme.}
\label{figgal}
\end{figure}

In order to test the Bose-Einstein Condensate dark matter model in the presence of random fluctuations, and with the effects of the baryonic matter fully taken into account, we have randomly selected from the SPARC database a set of 37 galaxies, from which 10 also contain information about their bulge velocity. For the other 27 galaxies this information is missing. The distribution of the considered set of galaxies according to their morphological classification scheme is presented in Fig.~\ref{figgal}. A significant number of galaxies (around 1/3) belong to the group of intermediate spiral galaxies $SAb$, that is, galaxies that are in between a barred spiral galaxy, and an unbarred spiral galaxy. The other considered galaxies have a relatively homogeneous distribution in our sample.

To fit Eq.~(\ref{69}), giving the velocity of a massive test particle in a Bose-Einstein Condensate in the presence of confining gravitational and random potentials we have fit the user-defined function (\ref{69}) to a set of data points $(x,y)$, or $(x,y,z)$ (the SPARC data), by using an implementation of the nonlinear least-squares (NLLS) Marquardt-Levenberg algorithm in Gnuplot. To obtain the optimal parameters the Chi-Square Statistic was applied, with $\chi ^2$ defined as
\be
\chi^2=\frac{1}{n-1}\sum_{i=1}^n{\frac{\left(O_i-E_i\right)^2}{E_i}},
\ee
where $n$ is the number of data points, $O_i$ are the observed values, while $E_i$ represent the expected values. We consider the fitting as good if $\chi^2<1$.

\subsection{Fitting results}

In the following we consider independently the two group of galaxies with and without bulge velocity information.  The number of fitting parameters is different in the two cases.

\subsubsection{The bulgeless galaxies}

In our study we considered 27 galaxies belonging to the bulgeless galaxies class.  The results of the comparison of the theoretical model (\ref{69}) with the SPARC observations are presented in Table~\ref{tableg1}. For the selected sample of galaxies we have obtained the central densities of the dark and baryonic matter $\left(\rho _c,\rho_{bc}\right)$, the static radius $R$, the rotation frequency $\omega$, the parameter $\gamma$ characterizing the baryonic matter distribution, the parameters $\left(\sigma, \theta\right)$ describing the effects of the random potential, as well as the stellar mass-to-light ratio for the disk $ \Upsilon_{d} $.

\begin{widetext}
\begin{center}
\begin{table}[htbp]
\begin{tabular}{|c|c|c|c|c|c|c|c|c|c|}
\hline
${\rm Galaxy}$ & $R$ (kpc) & ${\rho}_{c} \left(10^{-24} {\rm g/cm^{3}}\right)$ & $\rho_{bc} \left(10^{-24} {\rm g/cm^{3}}\right)$ & $\omega \;(10^{-16} {\rm s}^{-1})$ & $\sigma $  & $ \theta $ & $ \gamma $ & $ \Upsilon_{d} $ & $ \chi^{2} $ \\
\hline
DDO161 & 14.353 & 0.2060 & 0.01500 & 0.4070 & 0.01230 & 0.00139 & 1.9504 & 0.8408 & 0.293 \\
\hline
ESO079 & 16.810 & 0.8059 & 0.20600 & 1.5105 & 0.01080 & 0.00410 & 1.7739 & 1.2070 & 1.108 \\
\hline
ESO116& 10.350 & 0.9870 & 0.19740 & 1.8740 & 0.02060 & 0.00103 & 2.0940 & 0.7590 & 0.859 \\
\hline
F568-3& 19.470 & 0.3410 & 0.06820 & 0.7940 & 0.01900 & 0.00200 & 1.7450 & 0.8012 & 2.028  \\
\hline
F574-1& 13.610 & 0.5310 & 0.04620 & 1.1270 & 0.02670 & 0.00370 & 1.8450 & 0.8520 & 2.178 \\
\hline
NGC0055& 14.180 & 0.3810 & 0.07228 & 0.6740 & 0.01011 & 0.00127 & 1.7535 & 0.2881 & 0.737 \\
\hline
NGC0100 & 10.690 & 0.6080 & 0.12160 & 0.9270 & 0.01100 & 0.00120 & 1.4335 & 0.8040 & 0.225 \\
\hline
NGC0247 & 15.950 & 0.4140 & 0.08280 & 0.8186 & 0.02065 & 0.00120 & 2.0170 & 0.7067 & 1.988 \\
\hline
NGC2976 & 3.180 & 6.6810 & 1.27620 & 1.7040 & 0.02900 & 0.00350 & 2.1080 & 0.2912 & 0.389 \\
\hline
NGC3109 & 7.167 & 0.7230 & 0.14460 & 1.1230 & 0.01200 & 0.00113 & 1.9450 & 2.3200 & 0.193 \\
\hline
NGC3741 & 8.120 & 0.5217 & 0.10430 & 0.9120 & 0.01000 & 0.00125 & 2.0904 & 0.9102 & 2.745 \\
\hline
NGC3917 & 16.410 & 0.7470 & 0.14940 & 1.1030 & 0.01020 & 0.00410 & 2.0840 & 0.4008 & 2.876 \\
\hline
NGC3949& 7.860 & 3.1810 & 0.63620 & 1.4040 & 0.02510 & 0.00280 & 2.1450 & 0.3692 & 0.573 \\
\hline
NGC3972 & 9.690 & 1.7810 & 0.35620 & 1.4210 & 0.01080 & 0.00110 & 1.9450 & 0.4510 & 1.241 \\
\hline
NGC4010 & 11.110 & 1.2610 & 0.25220 & 0.3140 & 0.01600 & 0.00115 & 1.9240 & 0.2930 & 0.874 \\
\hline
NGC4183 & 23.350 & 0.2010 & 0.04020 & 0.3100 & 0.01100 &0.00120 & 1.7240 & 1.2592 & 1.217 \\
\hline
NGC4559 & 21.930 & 0.2510 & 0.05020 & 0.4760 & 0.02650 & 0.00112 & 1.6920 & 0.6800 & 0.135 \\
\hline
NGC5585 & 11.920 & 0.6070 & 0.13740 & 0.9230 & 0.02500 & 0.00250 & 1.8194 & 0.5802 & 1.613 \\
\hline
UGC00128 & 59.020 & 0.0720 & 0.01040 & 0.3450 & 0.02630 & 0.00100 & 1.4240 & 1.4502 & 2.988 \\
\hline
UGC00731 & 12.120 & 0.4310 & 0.05070 & 0.4010 & 0.01000 & 0.00120 & 1.7240 & 1.1102 & 2.657 \\
\hline
UGC05716 & 13.740 & 0.3060 & 0.06120 & 0.7830 & 0.02600 & 0.00410 & 1.8850 & 2.1730 & 0.978 \\
\hline
UGC05750 & 25.180 & 0.1157 & 0.02314 & 0.5150 & 0.02070 & 0.00128 & 1.9894 & 0.3318 & 1.727 \\
\hline
UGC06917 & 11.530 & 0.8723 & 0.17420 & 0.7198 & 0.01400 & 0.00280 & 1.9140 & 0.7502 & 0.687 \\
\hline
UGC07524 & 11.270 & 0.4900 & 0.08160 & 0.8920 & 0.01840 & 0.00320 & 2.2530 & 1.5960 & 0.685 \\
\hline
UGC09037 & 31.260 & 0.2491 & 0.04910 & 0.7810 & 0.01050 & 0.00250 & 2.1240 & 0.3102 & 1.483 \\
\hline
UGC12632 & 11.340 & 0.2060 & 0.04120 & 1.1980 & 0.03000 & 0.00120 & 1.9840 & 2.5210 & 0.594 \\
\hline
UGC12732 & 17.110 & 0.2450 & 0.04900 & 0.9080 & 0.02700 & 0.00120 & 1.8840 & 2.1202 & 0.773 \\
\hline
\end{tabular}
\caption{Optimal parameters for 27 bulgeless SPARC galaxies. In the Table $R$ denotes the static radius of the condensate, $\rho_c$ is the central density of the dark matter, $\rho_{bc}$ is the central density of the baryonic matter, $\omega$ is the angular rotational velocity of the halo, $\sigma$ and $\theta$ characterize the properties of the random disorder potential $\rho_{dis}=\rho_c \theta e^{-\sigma \xi^2}$,  $\gamma$ describes the baryonic matter density profile $\rho_b=\rho_{bc}e^{-\gamma \xi}$, while $\Upsilon _d$ is the mass-to-light ratio of the galactic disk.}\label{tableg1}
\end{table}
\end{center}
\end{widetext}

The physical parameters of the galaxies show relatively large variations. Thus, for example, the static radius of the galaxies varies in the range $(3.18,31.60)$ kpc. Similar variations do appear for the central densities of the dark matter and baryonic matter distributions. However, the angular velocity of the galaxy shows a relative constancy, with values of the order of $10^{-16}$ s$^{-1}$. Also the parameters $(\sigma, \theta, \gamma)$ have relatively constant values for the considered range of galaxies, and their fluctuations with respect to their mean value is small. The stellar mass-to-light ratio for the disk is generally smaller than one, with a few exceptions when it exceeds 2. We would like to mention that no restrictions on this quantity have been imposed when performing the fitting.

The results of the fits of the rotation curves are displayed in Fig.~\ref{rotg1}. In the Figures we have represented the observational data $v_{obs}$, the theoretical prediction due the condensate dark matter component $v_{DM}$, the baryonic component $v_{bar}$, obtained from the SPARC data, as well as $v_{tot}$ obtained from these quantities. $v_{tot}$ is then plotted, and compared to $v_{obs}$.  Generally, the theoretical model of the Bose-Einstain Condensate dark matter model in the presence of the baryonic matter and random interactions gives a good description of the observational data. This can be also inferred from the numerical values of $\chi^2$, which can also be seen in Table~\ref{tableg1}, with values generally smaller than one, with a few exceptions where $\chi ^2$ is of the order of 1.2, and a single example of a galaxy with $\chi ^2=2.704$. Higher values of $\chi^2$ are also correlated with extremely high values of $R$, indicating that for these galaxies the Bose-Einstein Condensate description of dark matter may be not particularly adequate.

\begin{widetext}
\begin{center}
\begin{figure*}[htbp!]
\centering
\includegraphics[width=11cm]{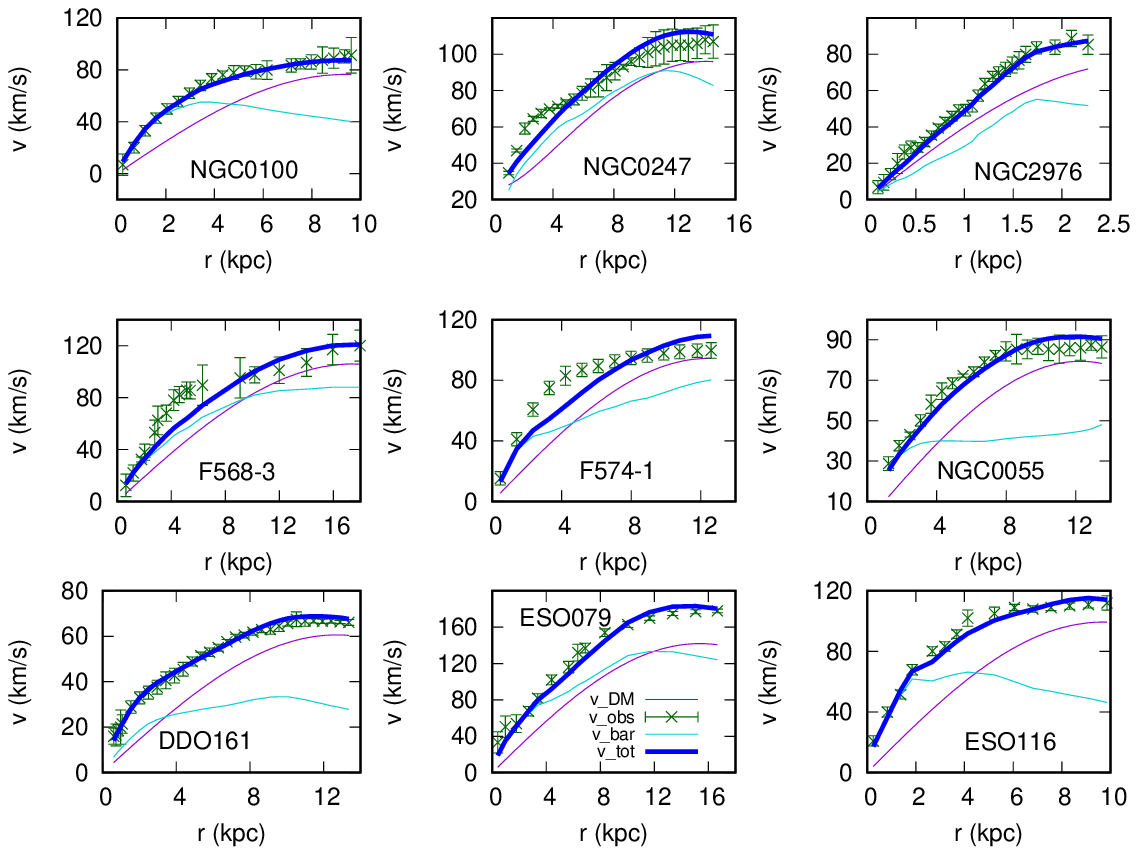}
\includegraphics[width=11cm]{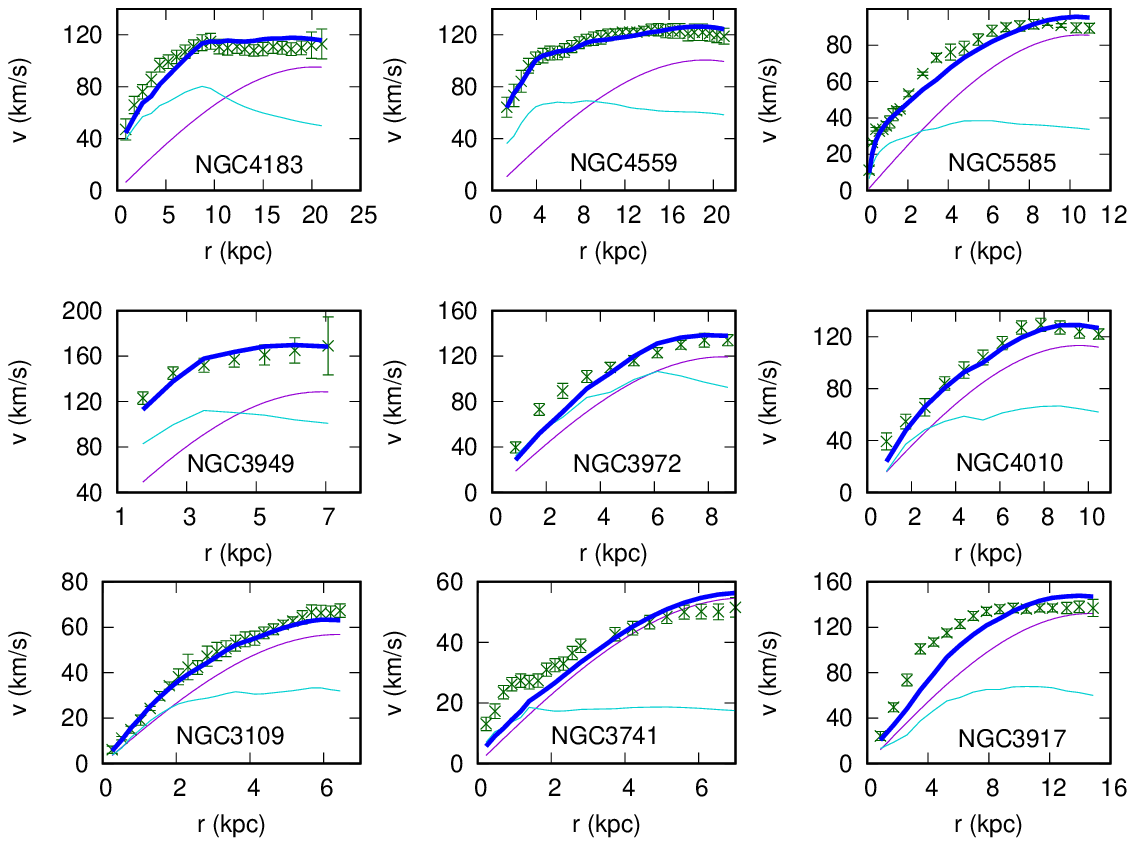}
\includegraphics[width=11cm]{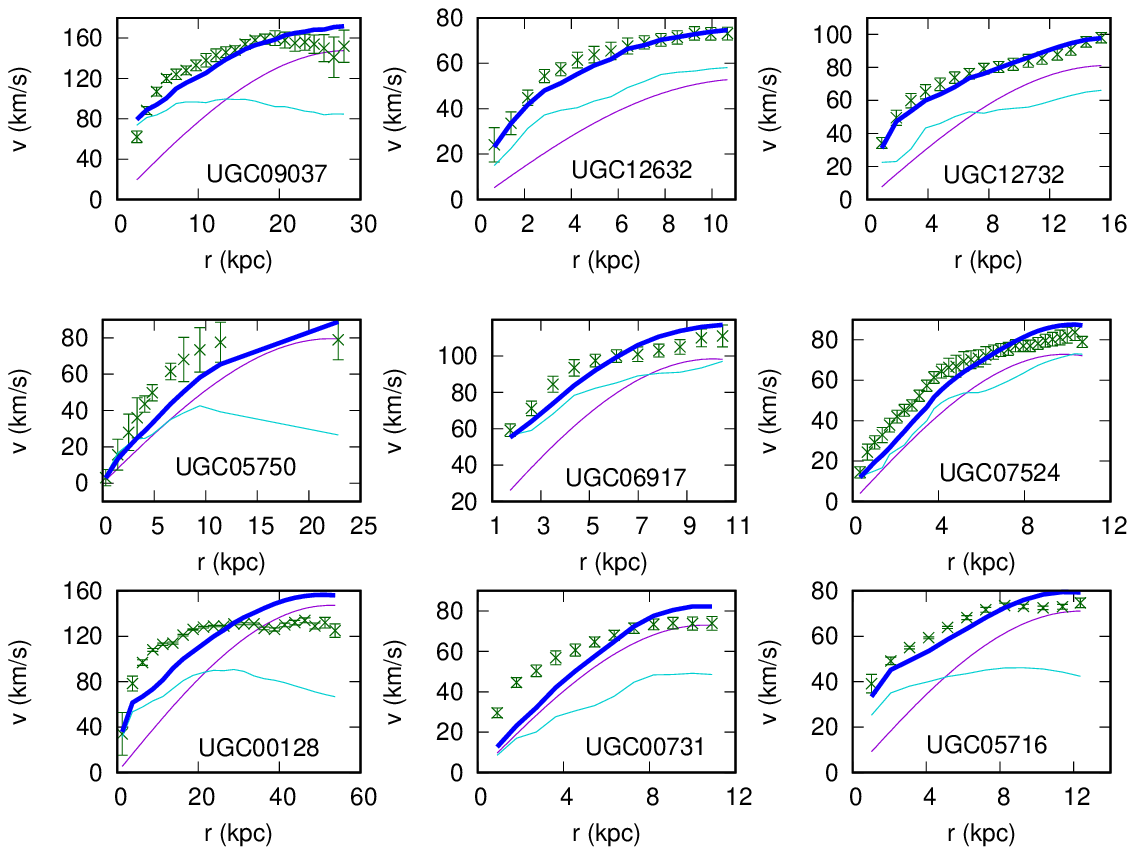}
\caption{Comparison of the theoretical predictions for the galactic rotation of the random Bose-Einstein Condensate dark matter model with 27 SPARC galaxies without bulge velocity. In the Figures  $v_{obs}$,  $v_{bar}$, $v_{DM}$ and $v_{tot}$ are represented by the dark - green, the dark - turquoise, the dark - violet, and the blue curves, respectively.}\label{rotg1}
\end{figure*}
\end{center}
\end{widetext}

\subsubsection{Fitting of the galaxies with bulge velocity}

The results of the fit of the 12 SPARC galaxies with bulge velocity are shown in Table~\ref{tableg2}. In this case we have considered one more astrophysical parameter, the  stellar mass-to-light ratios for the stellar bulge $\Upsilon_{b}$, which quantifies the contribution of the bulge velocity to the observed values of the rotational velocity. For these classes of galaxies we can observe a significant effect on the static radius $R$, which notably increases as compared with the bulgeless case. The other physical parameters do not show similar variations.

\begin{widetext}
\begin{center}
\begin{table}[htbp]
\begin{tabular}{|c|c|c|c|c|c|c|c|c|c|c|}
\hline
$Galaxy$ & $R $ (kpc)& ${\rho}_{c} \left(10^{-24} {\rm g/cm^{3}}\right)$ & ${\rho}_{bc} \left(10^{-24} {\rm g/cm^{3}}\right)$ & $\omega \;(10^{-16} {\rm s}^{-1})$ & $\sigma $ & $ \theta $ & $ \gamma $ & $ \Upsilon_{d} $ & $ \Upsilon_{b} $ & $ \chi^{2} $ \\
\hline
IC4202 & 27.84 & 0.112 & 0.4520 & 0.898 & 0.0190 & 0.00230 & 1.9900 & 1.441 & 0.219 & 2.910  \\
\hline
NGC2841 & 64.24 & 0.562 & 3.0550 & 0.298 & 0.0250 & 0.00430 & 1.7600 & 1.074 & 1.170 & 1.039 \\
\hline
NGC4013 & 31.75 & 0.275 & 0.0012 & 0.828 & 0.0160 & 0.00220 & 0.5106 & 0.116 & 2.281 & 2.343 \\
\hline
NGC4157 & 30.45 & 0.243 & 0.0486 & 1.215 & 0.0160 & 0.00220 & 1.9700 & 0.547 & 0.098 & 0.208 \\
\hline
NGC5005 & 12.44 & 3.785 & 0.7010 & 0.828 & 0.0100 & 0.00100 & 1.4520 & 0.366 & 0.612 & 0.217 \\
\hline
NGC5985 & 35.24 & 0.180 & 0.0091 & 1.354 & 0.0217 & 0.00325 & 0.2030 & 1.712 & 1.797 & 1.931  \\
\hline
NGC6195 & 37.57 & 0.345 & 0.0027 & 0.487 & 0.0185 & 0.00270 & 0.7050 & 0.405 & 0.649 & 2.104 \\
\hline
NGC7814 & 20.64 & 0.672 & 0.1544 & 1.624 & 0.0135 & 0.00170 & 1.9200 & 2.183 & 0.548 & 0.738 \\
\hline
UGC02885 & 74.48 & 0.115 & 0.0109 & 0.715 & 0.0270 & 0.00440 & 2.1900 & 0.938 & 1.025 & 1.391 \\
\hline
UGC06614 & 65.45 & 0.157 & 0.0654 & 0.806 & 0.0259 & 0.00435 & 0.1580 & 0.895 & 0.577 & 0.498 \\
\hline
UGC06787 & 6.05 & 10.445 & 0.2710 & 1.328 & 0.0090 & 0.00090 & 1.8490 & 0.092 & 0.621 & 1.327 \\
\hline
UGC08699 & 26.48 & 0.315 & 0.0650 & 1.149 & 0.0136 & 0.00185 & 1.9520 & 1.335 & 0.552 & 0.284 \\
\hline
\end{tabular}
\caption{Optimal fits for 12 galaxies from the SPARC database with bulge velocities.}\label{tableg2}
\end{table}
\end{center}
\end{widetext}

The comparison of the theoretical predictions for the rotation velocity curves of the 12 SPARC galaxies with bulge velocities,  and observations is presented Fig.~\ref{rotg2}.

\begin{widetext}
\begin{center}
\begin{figure}[h!]
\centering
\includegraphics[width=11cm]{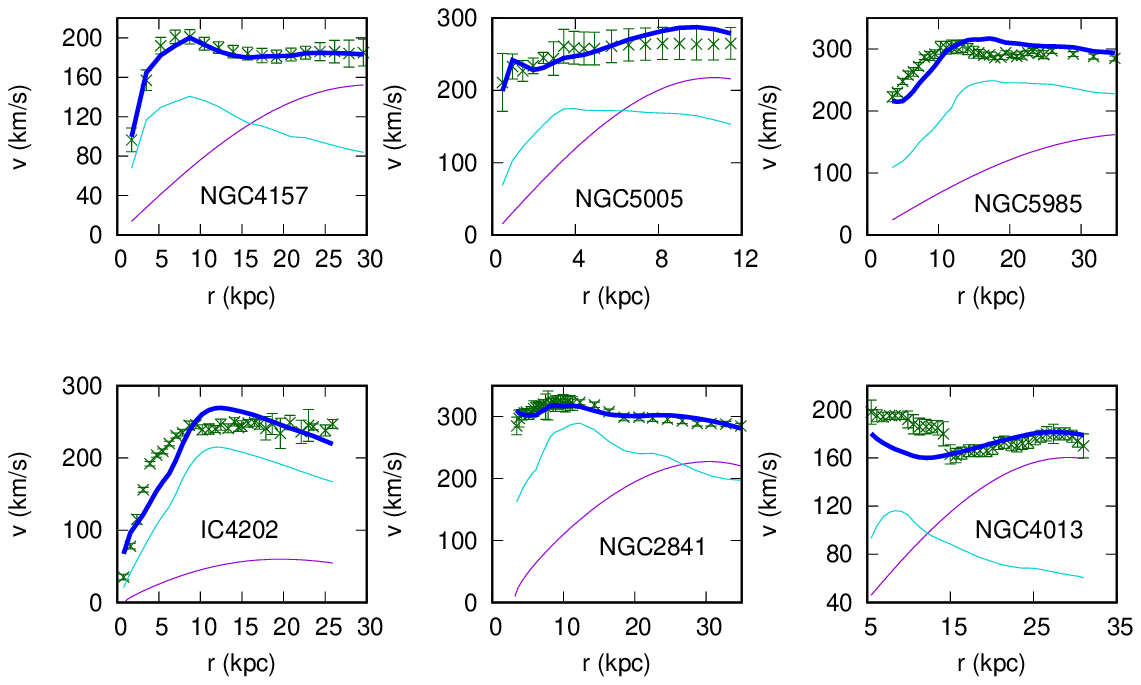}
\includegraphics[width=11cm]{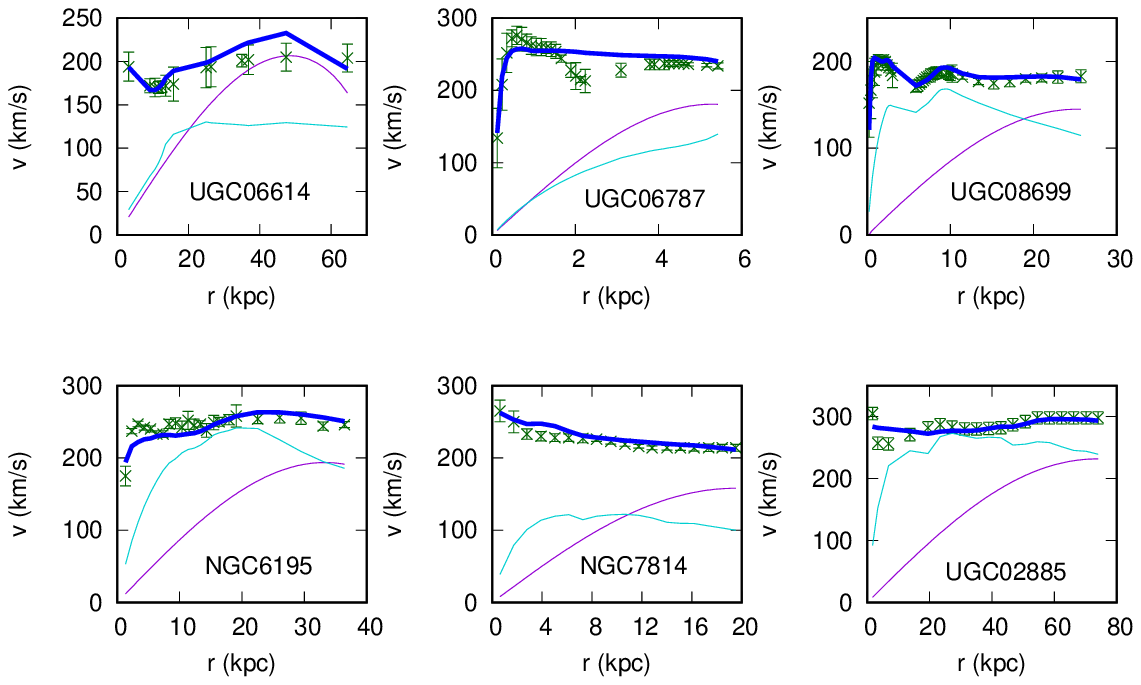}
\caption{Comparison of the theoretical predictions for the galactic rotation of the random Bose-Einstein Condensate dark matter model with 12 SPARC galaxies with bulge velocity.}
\label{rotg2}
\end{figure}
\end{center}
\end{widetext}

\subsubsection{Statistical results}

 As one can see from the fitting results presented above, the Bose-Einstein Condensate dark matter model also gives, under the assumption of a varying $R$, an acceptable description of the considered types of galaxies. The distribution of the $\chi^2$ values versus the number of galaxies for the entire considered set, including galaxies with and without bulge is presented in Fig.~\ref{fn1}.
 \begin{widetext}
 \begin{center}
\begin{figure*}[h!]
\centering
\includegraphics[width=8.5cm]{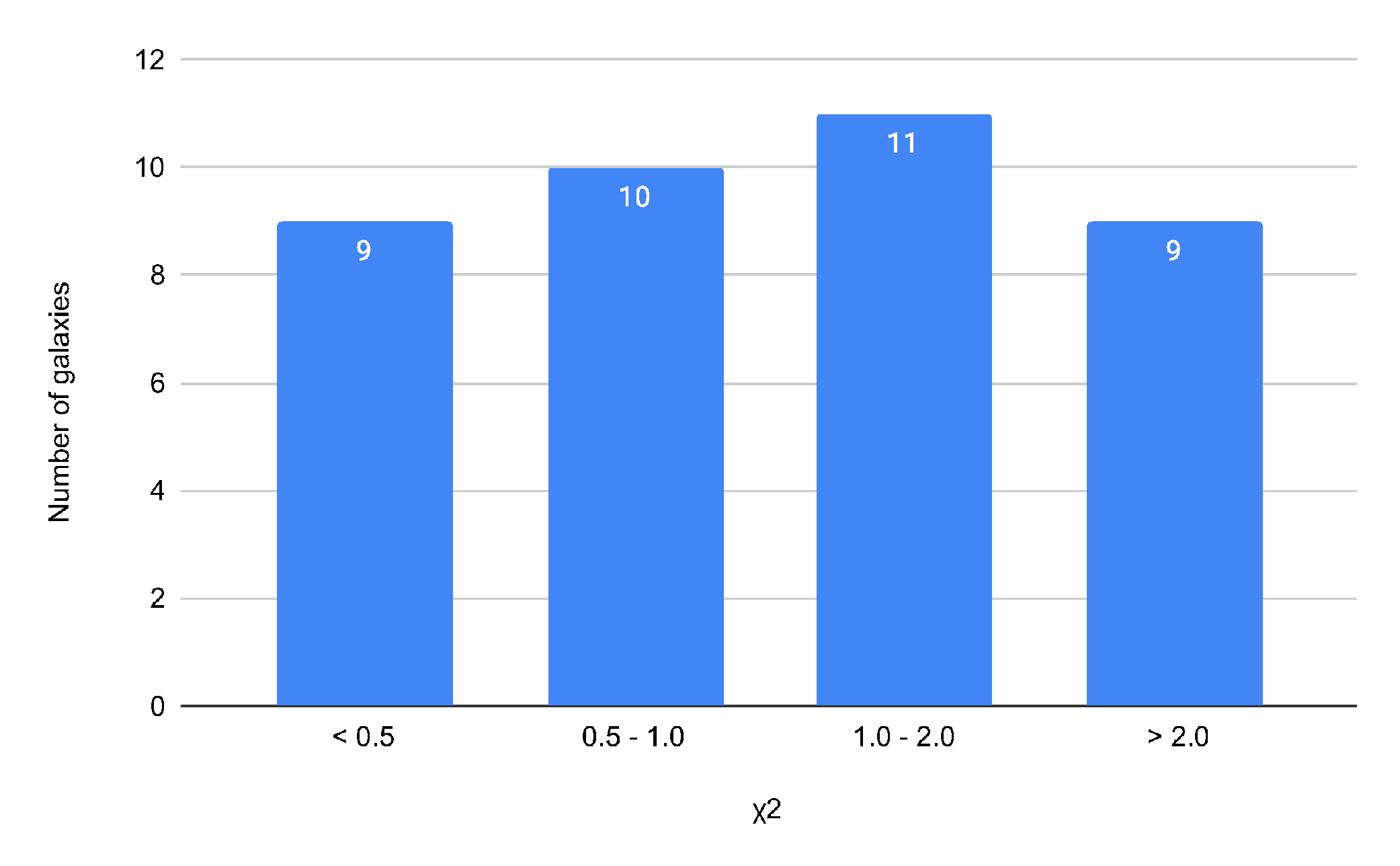}
\includegraphics[width=8.5cm]{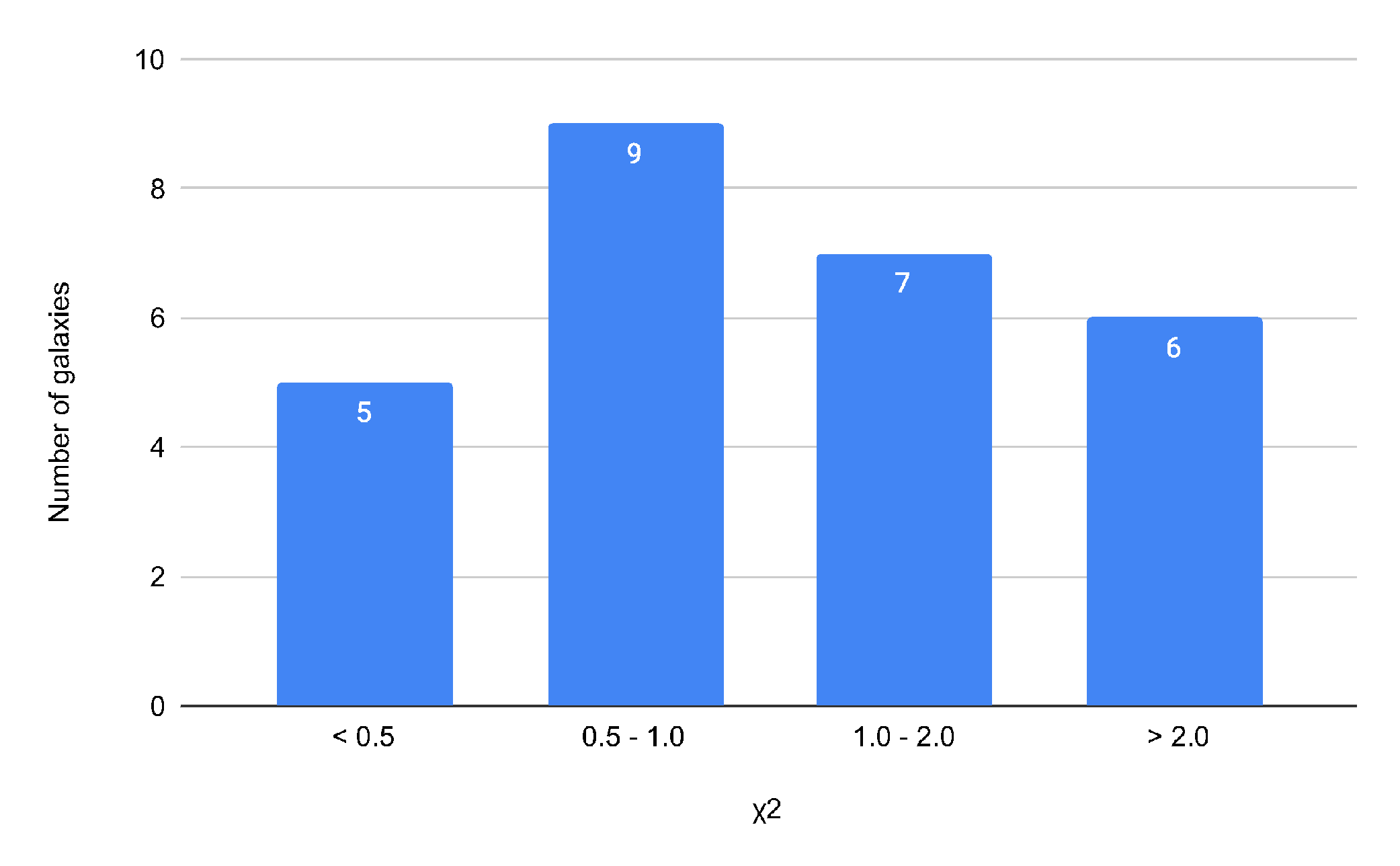}
\caption{The distribution of the $\chi^2$ values, describing the accuracy of the fitting of the rotation curves for the full set of 39 considered galaxies (left panel), and for the 27 bulgeless galaxies (right panel).}
\label{fn1}
\end{figure*}
\end{center}
\end{widetext}

 In the case of the total sample, for 19 galaxies, that is, around half of the sample, $\chi^2<1$, indicating very good fits. For the rest of the galaxies, 11 have $\chi^2$ values in the range of 1-2, indicating acceptable fits, while for 9 galaxies (around 25\% of the sample), $\chi^2>2$. In the case of the 27 bulgeless galaxies, 14 galaxies have a $\chi^2<1$, and hence very good fits, for 7 galaxies $\chi^2$ is in the range 1-2, with the fits considered acceptable, while for 6 galaxies $\chi^2>2$.
 For the considered set of 12 bulgeless galaxies, six have a $\chi ^2$ value smaller than one, while five have $\chi ^2$ values in the range $1<\chi^2<2$. For a single galaxy $\chi ^2$ is greater than 3.

\begin{center}
\begin{figure}[h!]
\centering
\includegraphics[width=8.5cm]{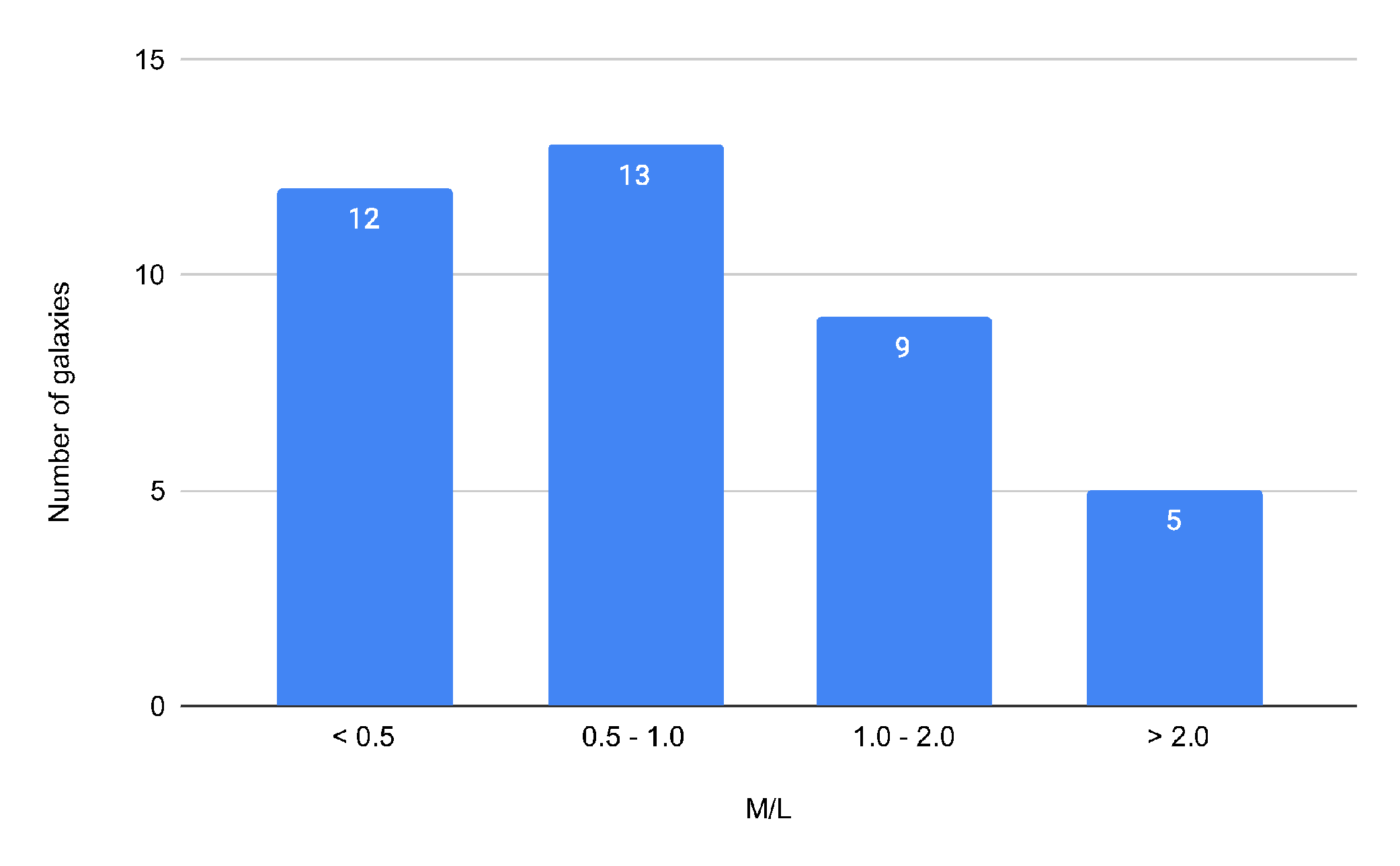}
\caption{ Distribution of the mass-to-light ratio values $\Upsilon_d$ for the disk  for the set of the 27 considered bulgeless galaxies}.
\label{fnML1}
\end{figure}
\end{center}

 The fits of the rotation curves have been performed by leaving the mass-to-light ratios for the disk and for the stellar bulge unconstrained. A value of the mass-to-light ratio of around 0.5 is in good agreement is in good agreement with predictions from stellar population synthesis models \cite{SchM}. The statistical distribution of the mass-to-light ratio $\Upsilon_d$ for the disk is represented, in terms of the considered galaxy numbers, in Fig.~\ref{fnML1}. The average value of $\Upsilon_d$  for the considered sample of galaxies is $\left<\Upsilon_d\right>=0.9695$, which differs from the standard 0.5 value. However, 66\% of the considered galaxies have mass-to-light ratios around smaller than one, while for 33\% this ratio is bigger. Similar results have been obtained for some particular galaxies in other studies. For example, for the galaxy NGC2841 the values $\Upsilon _d=0.81\pm 0.05$ and $\Upsilon_b=0.93\pm 0.05$ have been obtained in \cite{Lin}. On the other hand, in the same study, for the galaxy IC2574 the value $\Upsilon_d=0.07\pm 0.01$ was found.

  Hence, the consistency of the $M/L$ values with the astrophysical observations may provide a powerful test of the Bose-Einstein Condensate theoretical models, and of their viability. Recently, an investigation of the possibility of fitting  the galactic rotation curves with reasonable stellar mass-to-light ratios was performed in \cite{Mist}, by fitting the superfluid dark matter model to the rotation curves of 169 galaxies in the SPARC sample. In this study it was found that the mass-to-light ratios obtained with superfluid dark matter are consistent in terms of stellar populations. But it was also found  that the best fit mass-to-light ratios depend on the size of the galaxy, with giant galaxies having systematically lower mass-to-light ratios than dwarf galaxies.

\subsection{Checking the constancy of $R$}

One of the fundamental predictions of the Bose-Einstein Condensate dark matter model is the existence of a quantity $R$, having the physical/astrophysical  meaning as the radius of the galactic halo, and which is a Universal constant, determined by two basic properties of dark matter, the mass and scattering length of its component particle. However, when performing a general fit of the rotation velocity curves with Bose-Einstein Condensate type dark matter models, values of $R$ in a large range are generated, a result which naturally raises the question of the validity of the theoretical model. In the following we investigate this problem from the point of view of the observations, and we redo the fittings by considering  that the static radius of the condensate $R$ is {\it a universal constant}. In order to perform the fitting we need to fix the numerical value of $R$. Since the mass of the dark matter particle as well as its scattering length are presently unknown, we adopt for $R$ a somehow smaller value than the average value obtained from the general fitting of the 27 bulgeless galaxies, as presented in Table~\ref{tableg1}. Hence, in order to also optimize the fits, we investigate the behavior of the galactic rotation curves in the random Bose-Einstein Condensate model with constant $R$ by adopting the value $R=16$ kpc.

\subsubsection{Fitting results for bulgeless galaxies with constant $R$.}

 The results of the fitting of the 27 SPARC galaxies under the assumption of a constant $R$ are presented in Table~\ref{tableg3a}. There are no significant changes in the overall results, the theoretical model with constant $R$ also giving an excellent fit to the data, with most of the $\chi ^2$ values below one, and generally unchanged. Hence, the interpretation of the observational data does not rule out a constant value of $R$. Of course, finding its optimal value requires deeper investigations into the theoretical models, and the determination of the physical parameters that play the dominant role in the model.

 \begin{widetext}
\begin{center}
\begin{figure*}[htbp]
\centering
\includegraphics[width=11cm]{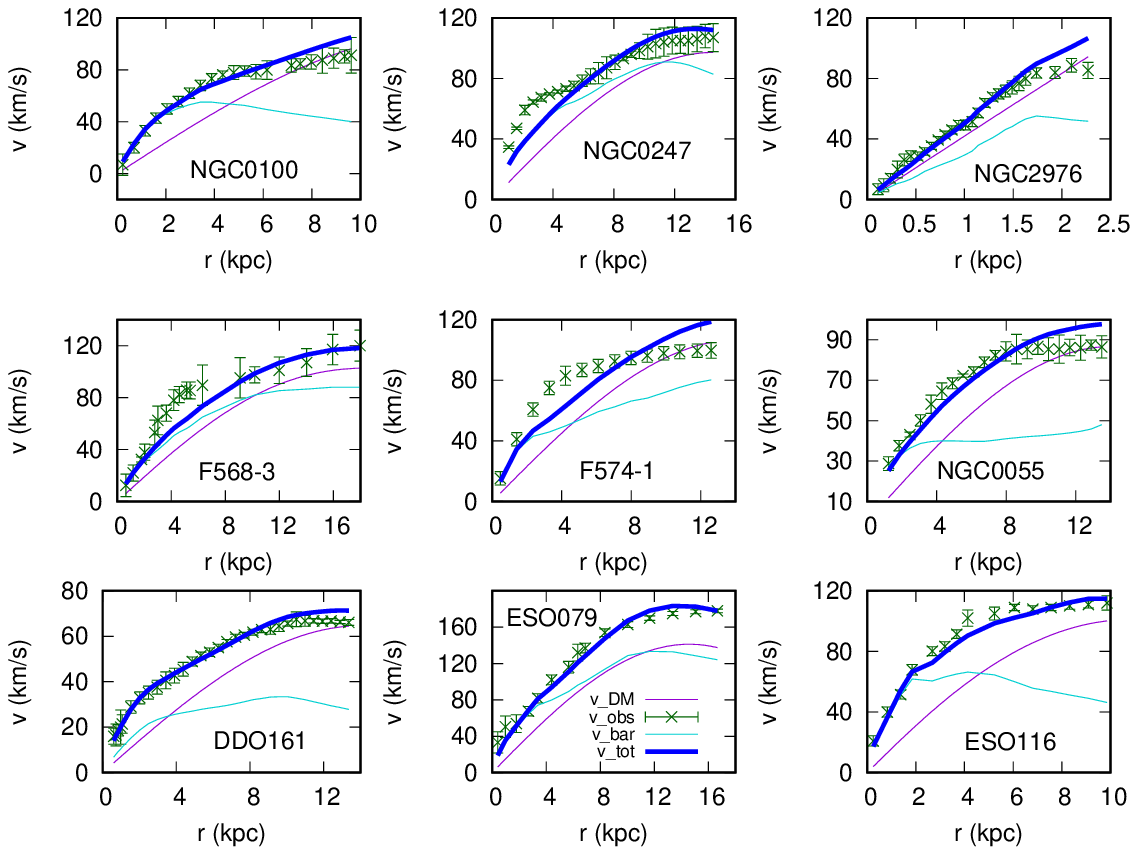}
\includegraphics[width=11cm]{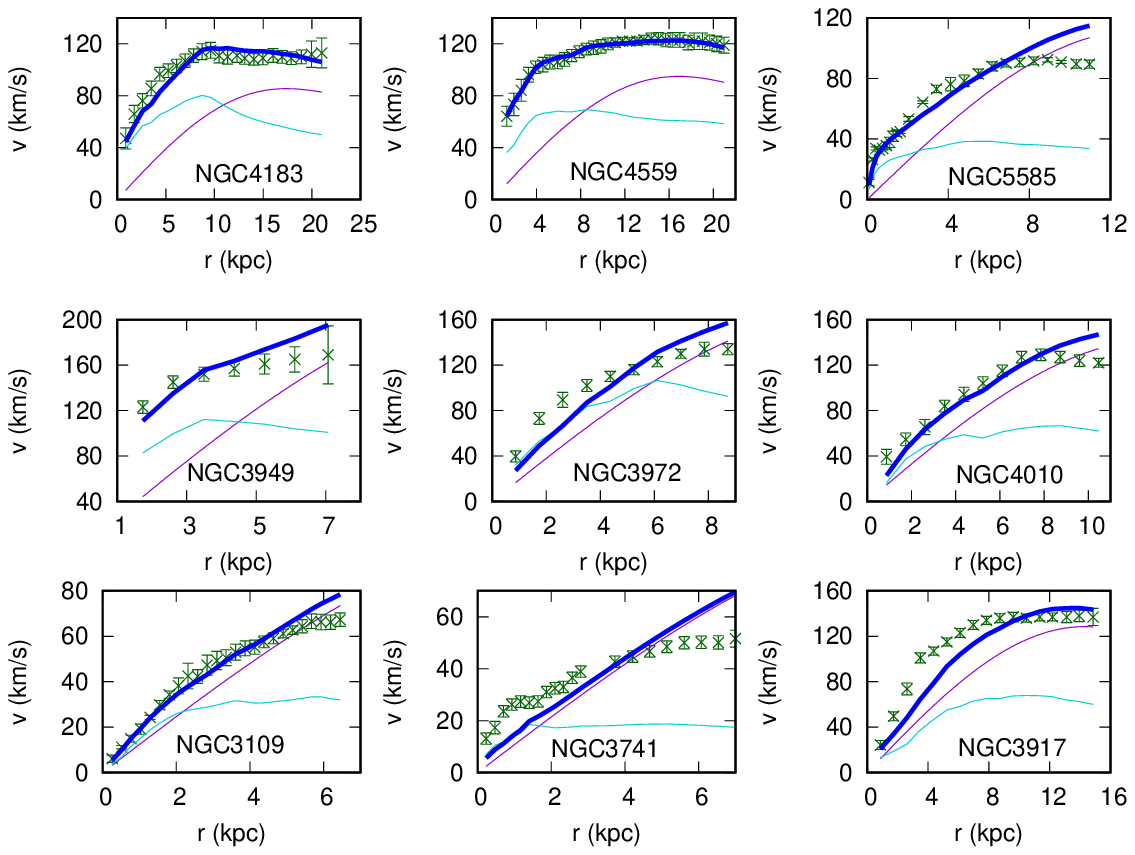}
\includegraphics[width=11cm]{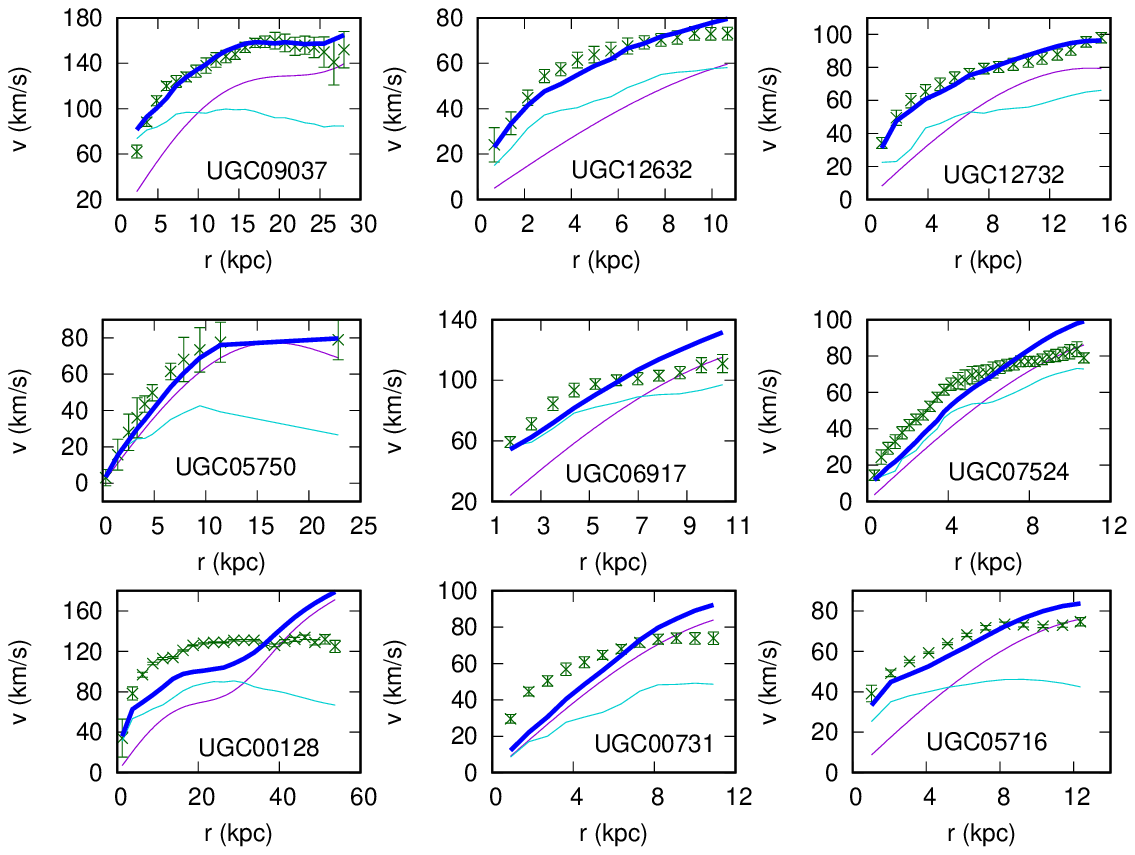}
\caption{Comparison of the theoretical predictions for the galactic rotation of the random Bose-Einstein Condensate dark matter model with 27 SPARC galaxies without bulge velocity for a universal static radius $R$.}
\label{rotg3}
\end{figure*}
\end{center}
\end{widetext}

\begin{widetext}
\begin{center}
\begin{table}[htbp]
\begin{tabular}{|c|c|c|c|c|c|c|c|c|c|}
\hline
${\rm Galaxy}$ & $R$ (kpc) & ${\rho}_{c} \left(10^{-24} {\rm g/cm}^{3}\right)$ & $\rho_{bc}\left(10^{-24} {\rm g/cm}^{3}\right)$ & $\omega \;(10^{-16} {\rm s}^{-1})$ & $\sigma $  & $ \theta $ & $ \gamma $ & $ \Upsilon_{d} $ & $ \chi^{2} $ \\
\hline
DDO161 & 16.0 & 0.1925 & 0.02410 & 0.3470 & 0.02210 & 0.00198 & 1.8250 & 0.8408 & 1.034 \\
\hline
ESO079 & 16.0 & 0.8759 & 0.17518 & 1.5105 & 0.01970 & 0.00272 & 1.7340 & 0.6868 & 0.987 \\
\hline
ESO116& 16.0 & 0.9570 & 3.29740 & 1.8740 & 0.02310 & 0.00189 & 1.9170 & 1.2412 & 0.893 \\
\hline
F568-3& 16.0 & 0.3410 & 0.06820 & 1.7940 & 0.01900 & 0.00200 & 1.7450 & 0.8012 & 2.032  \\
\hline
F574-1& 16.0 & 0.5120 & 0.03220 & 0.7270 & 0.02230 & 0.00210 & 1.6270 & 0.8524 & 2.217 \\
\hline
NGC0055& 16.0 & 0.3510 & 0.07228 & 0.5080 & 0.02240 & 0.00252 & 1.8380 & 0.3981 & 0.896 \\
\hline
NGC0100 & 16.0 & 0.5580 & 0.12160 & 0.8670 & 0.02310 & 0.00192 & 1.9790 & 0.7840 & 0.473 \\
\hline
NGC0247 & 16.0 & 0.4140 & 0.08280 & 1.0186 & 0.02250 & 0.00242 & 1.8920 & 0.8067 & 1.997 \\
\hline
NGC2976 & 16.0 & 6.6810 & 0.77620 & 0.9040 & 0.02900 & 0.00350 & 2.1080 & 0.2912 & 0.498\\
\hline
NGC3109 & 16.0 & 0.5830 & 0.11460 & 0.9230 & 0.02460 & 0.00128 & 1.9850 & 1.8270 & 0.278 \\
\hline
NGC3741 & 16.0 & 0.4417 & 0.10410 & 0.7120 & 0.02480 & 0.00127 & 1.9570 & 0.6502 & 2.983\\
\hline
NGC3917 & 16.0 & 0.7470 & 0.14940 & 1.1030 & 0.02320 & 0.00248 & 1.8550 & 0.4008 & 2.783 \\
\hline
NGC3949& 16.0 & 2.4810 & 0.62620 & 0.8040 & 0.02460 & 0.00172 & 1.7230 & 0.3492 & 0.705 \\
\hline
NGC3972 & 16.0 & 1.3810 & 0.36620 & 0.8210 & 0.02400 & 0.00129 & 1.9140 & 0.4510 & 1.578 \\
\hline
NGC4010 & 16.0 & 1.0410 & 0.23220 & 0.3840 & 0.02390 & 0.00192 & 0.9390 & 0.2530 & 0.936 \\
\hline
NGC4183 & 16.0 & 0.2510 & 0.04020 & 1.4170 & 0.01580 &0.00276 & 1.9870 & 1.5420 & 1.223 \\
\hline
NGC4559 & 16.0 & 0.3210 & 0.05020 & 1.5260 & 0.01570 & 0.00267 & 1.9740 & 0.6870 & 0.079 \\
\hline
NGC5585 & 16.0 & 0.6270 & 0.11740 & 0.6030 & 0.02380 & 0.00194 & 0.7350 & 0.5372 & 1.893 \\
\hline
UGC00128 & 16.0 & 0.1180 & 0.00940 & 1.2450 & 0.02980 & 0.00210 & 2.2450 & 2.4020 & 3.097 \\
\hline
UGC00731 & 16.0 & 0.3710 & 0.05070 & 0.3210 & 0.01800 & 0.00140 & 2.3030 & 2.2702 & 2.889 \\
\hline
UGC05716 & 16.0 & 0.2760 & 0.06120 & 0.6030 & 0.02900 & 0.00243 & 1.9230 & 2.3030 & 1.108 \\
\hline
UGC05750 & 16.0 & 0.2157 & 0.01814 & 1.2150 & 0.01950 & 0.00318 & 1.5810 & 0.3208 & 0.983 \\
\hline
UGC06917 & 16.0 & 0.7323 & 0.12420 & 0.5098 & 0.02280 & 0.00221 & 1.8980 & 0.8002 & 0.914 \\
\hline
UGC07524 & 16.0 & 0.3980 & 0.08160 & 0.7420 & 0.02350 & 0.00213 & 1.8020 & 0.8270 & 0.972 \\
\hline
UGC09037 & 16.0 & 0.4791 & 0.04910 & 2.2810 & 0.02280 & 0.00334 & 1.4830 & 0.3502 & 0.495 \\
\hline
UGC12632 & 16.0 & 0.1860 & 0.04120 & 0.7980 & 0.02380 & 0.00207 & 1.9170 & 2.5180 & 0.821 \\
\hline
UGC12732 & 16.0 & 0.2750 & 0.04900 & 0.8580 & 0.02580 & 0.00269 & 1.8120 & 2.2602 & 0.785 \\
\hline
\hline
\end{tabular}
\caption{Optimal fitting parameters for 27 bulgeless SPARC galaxies with constant static radius $R$.}\label{tableg3a}
\end{table}
\end{center}
\end{widetext}

The fitting of the rotation curves for the 27 bulgeless SPARC galaxies, under the assumption of a constant, universal static radius $R$ are represented in Fig.~\ref{rotg3}. Generally, the fits give a good description of the observational data, and the one-parameter reduced model gives equally good fits like the eight parameter case. The assumption of a constant $R$ does not introduce drastic differences in the description of the observational data by the theoretical model.

\subsubsection{Galaxies with bulge velocities and constant $R$.}

As one can see from Table~\ref{tableg2}, the results of the fitting of the 12 SPARC galaxies with bulge provided values of $R$ significantly different from those obtained for the bulgeless galaxies, with much higher values of the static radius. That's why it is of particular interest to investigate the possibility of the existence of a universal value of $R$, which is the same as for the bulgeless galaxies.  The results of the fitting of the theoretical model with constant $R$ with 12 SPARC galaxies with bulge velocity are presented in Table~\ref{tableg4}. Generally, for the adopted constant value of $R$, the model gives a good description of the observed rotational velocity curves, with the values of $\chi ^2$ not exceeding 2,42. For most galaxies the central density of the condensate increases slightly, but in the case of the galaxy UGC06787 it decreases from $11.445\times 10^{-24}$ g/cm$^3$ to $4.235\times 10^{-24}$ g/cm$^3$.

\begin{widetext}
\begin{center}
\begin{table}
\begin{tabular}{|c|c|c|c|c|c|c|c|c|c|c|c|c|}
\hline
${\rm Galaxy}$ & $R ({\rm kpc}) $ & ${\rho}_{c} \left(10^{-24} {\rm g/cm}^{3}\right)$ & ${\rho}_{bc} \left(10^{-24} {\rm g/cm}^{3}\right)$ & $\omega \;(10^{-16} {\rm s}^{-1})$ & $\sigma $ & $ \theta $ & $ \gamma $ & $ \Upsilon_{d} $ & $ \Upsilon_{b} $ & $ \chi^{2} $ \\
\hline
IC4202 & 16.0 & 0.524 & 0.8820 & 2.738 & 0.0210 & 0.00210 & 2.0700 & 1.071 & 0.245 & 2.797 \\
\hline
NGC2841 & 16.0 & 0.442 & 0.7550 & 3.098 & 0.0210 & 0.00380 & 1.9600 & 1.345 & 1.173 & 1.573 \\
\hline
NGC4013 & 16.0 & 0.204 & 0.0322 & 2.174 & 0.0100 & 0.00137 & 1.7800 & 1.076 & 0.191 & 2.179 \\
\hline
NGC4157 & 16.0 & 0.592 & 0.0613 & 2.423 & 0.0104 & 0.00143 & 1.7520 & 0.473 & 0.293 & 0.388 \\
\hline
NGC5005 & 16.0 & 1.724 & 0.7030 & 0.512 & 0.0160 & 0.00158 & 1.4590 & 0.489 & 0.608 & 0.108 \\
\hline
NGC5985 & 16.0 & 0.088 & 0.0102 & 1.398 & 0.0120 & 0.00183 & 1.8200 & 1.974 & 1.897 & 1.883  \\
\hline
NGC6195 & 16.0 & 0.424 & 0.0009 & 2.258 & 0.0098 & 0.00143 & 1.8500 & 0.384 & 0.668 & 2.197 \\
\hline
NGC7814 & 16.0 & 1.152 & 0.1632 & 2.177 & 0.0129 & 0.00163 & 1.7260 & 1.291 & 0.524 & 0.868 \\
\hline
UGC02885 & 16.0 & 0.286 & 0.0574 & 1.301 & 0.0072 & 0.00117 & 2.2700 & 1.217 & 0.925 & 1.895 \\
\hline
UGC06614 & 16.0 & 0.204 & 0.0145 & 1.678 & 0.0078 & 0.00131 & 2.1080 & 0.592 & 0.619 & 1.254 \\
\hline
UGC06787 & 16.0 & 3.950 & 0.3240 & 1.128 & 0.0290 & 0.00290 & 1.7420 & 0.182 & 0.625 & 1.394 \\
\hline
UGC08699 & 16.0 & 0.413 & 0.0834 & 2.569 & 0.0110 & 0.00164 & 1.7490 & 1.135 & 0.581 & 0.279 \\
\hline
\end{tabular}
\caption{Optimal parameters for galaxies with bulge velocity, and constant static radius $R$.}\label{tableg4}
\end{table}
\end{center}
\end{widetext}

The fits of the rotational velocity curves of the galaxies with bulge velocity with the random Bose-Einstein Condensate dark matter model are presented, for a constant value of the static radius $R$, in Fig.~\ref{rotg4}. With the exception of the galaxy UGC06614, the random Bose-Einstein Condensate dark matter model with constant $R$ gives equivalent results with those in which $R$ is allowed to vary freely. Hence, the same constant numerical value of $R$ gives a good description of the observational data for both galaxies with and without a bulge.

\begin{widetext}
\begin{center}
\begin{figure}[tbp!]
\centering
\includegraphics[width=11cm]{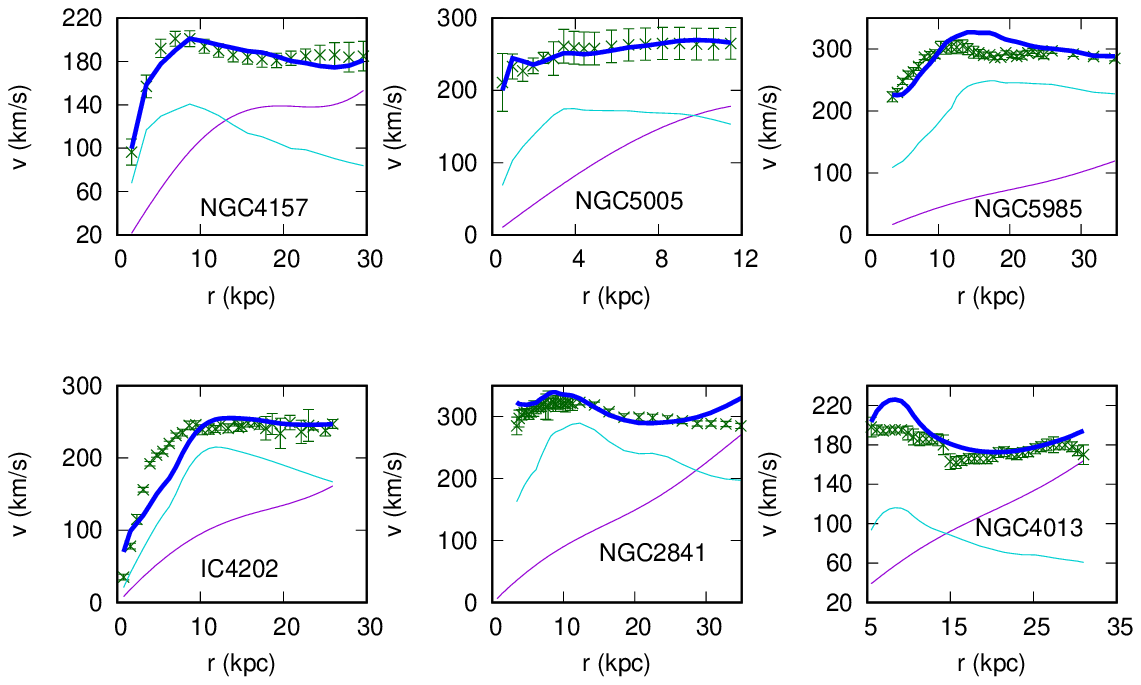}
\includegraphics[width=11cm]{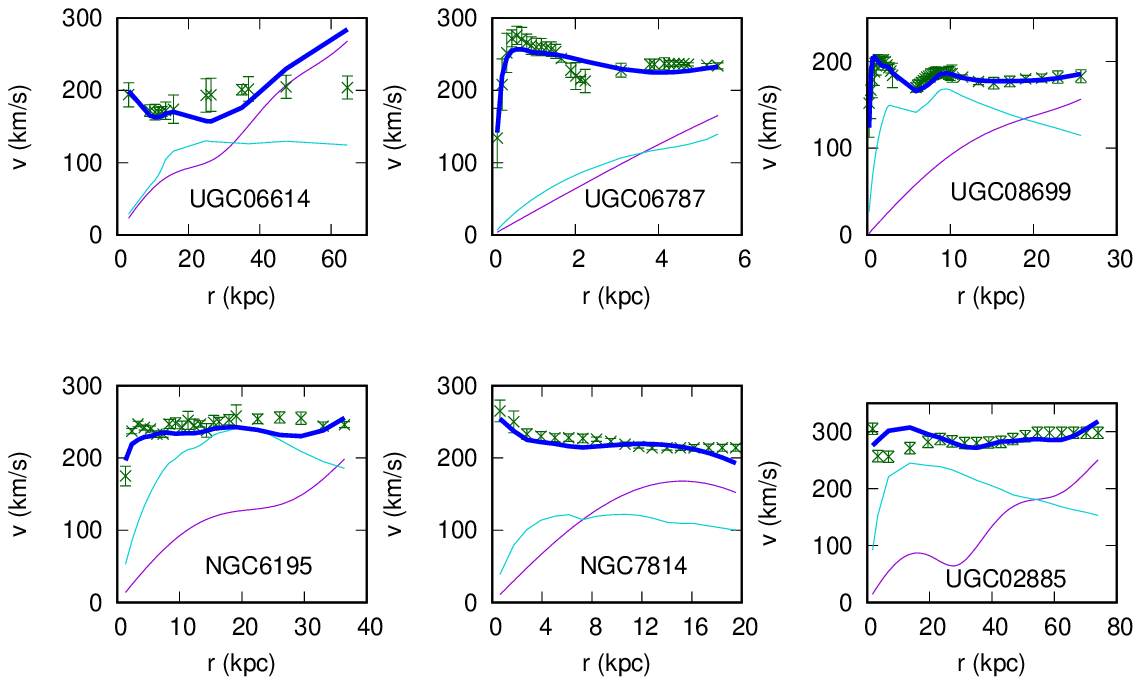}
\caption{Comparison of the theoretical predictions for the galactic rotation curves of the random Bose-Einstein Condensate dark matter model with constant static radius $R$ with the observational data of 12 SPARC galaxies with bulge velocity.}
\label{rotg4}
\end{figure}
\end{center}
\end{widetext}

\subsubsection{Analysis of the statistical results}

\begin{center}
\begin{figure}[h!]
\centering
\includegraphics[width=8.5cm]{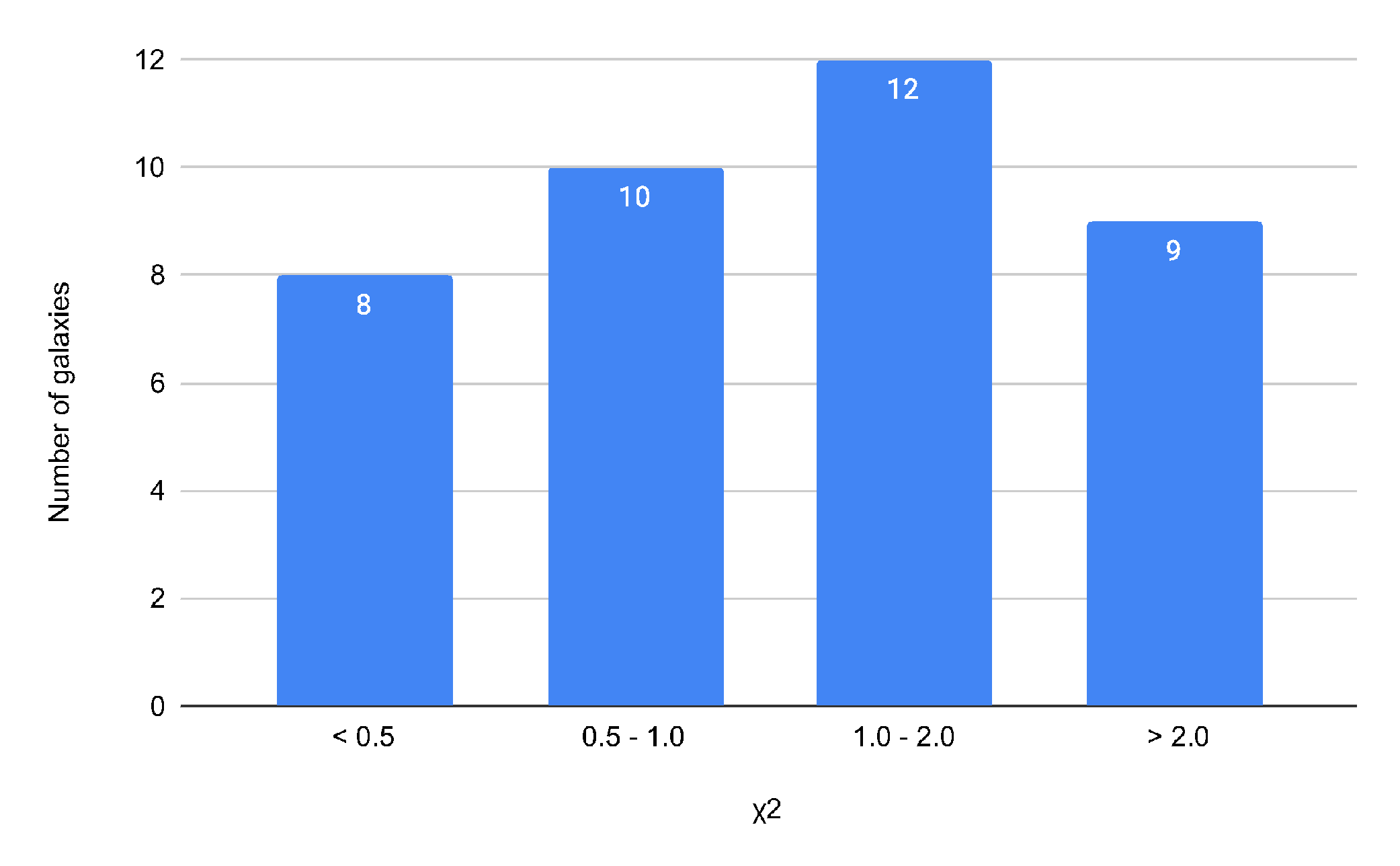}
\caption{The distribution of the $\chi^2$ values, describing the accuracy of the fitting of the rotation curves for the full set of 39 considered galaxies for a fixed static radius $R=16$ kpc.}
\label{fn2}
\end{figure}
\end{center}

The distribution of the $\chi^2$ values obtained from the fitting of the galactic rotation curves in the case of a fixed static radius $R=16 kpc$
is represented, for the full set of 39 considered galaxies, in Fig.~\ref{fn2}.

From a purely statistical point of view, there are no major differences between the varying $R$ and the fixed $R$ cases. Assuming a varying $R$, 19 galaxies have  $0.1<\chi^2<1$, while in the case of a fixed $R$ this number is of 18 galaxies. For arbitrary $R$, 11 galaxies have $1<\chi^2<2$, while the corresponding number of galaxies for $R$ fixed is 12. In both cases the number of galaxies with $\chi^2>2$ is 9. Hence, the present results indicate that if indeed dark matter is in the form of a Bose-Einstein Condensate, the possibility of the existence of a universal length scale $R$, describing the static properties of the condensed dark matter, and depending only on the mass and scattering length of teh dark matter particle, cannot be firmly ruled out by the observations. In the present study we have adopted, for the sake of simplicity, for $R$ an arbitrary value $R=16$ kpc. An interesting and important problem would be to find a systematic and consistent mathematical method to extract the exact value of $R$ from the astrophysical observations.

 \begin{center}
\begin{figure}[h!]
\centering
\includegraphics[width=8.5cm]{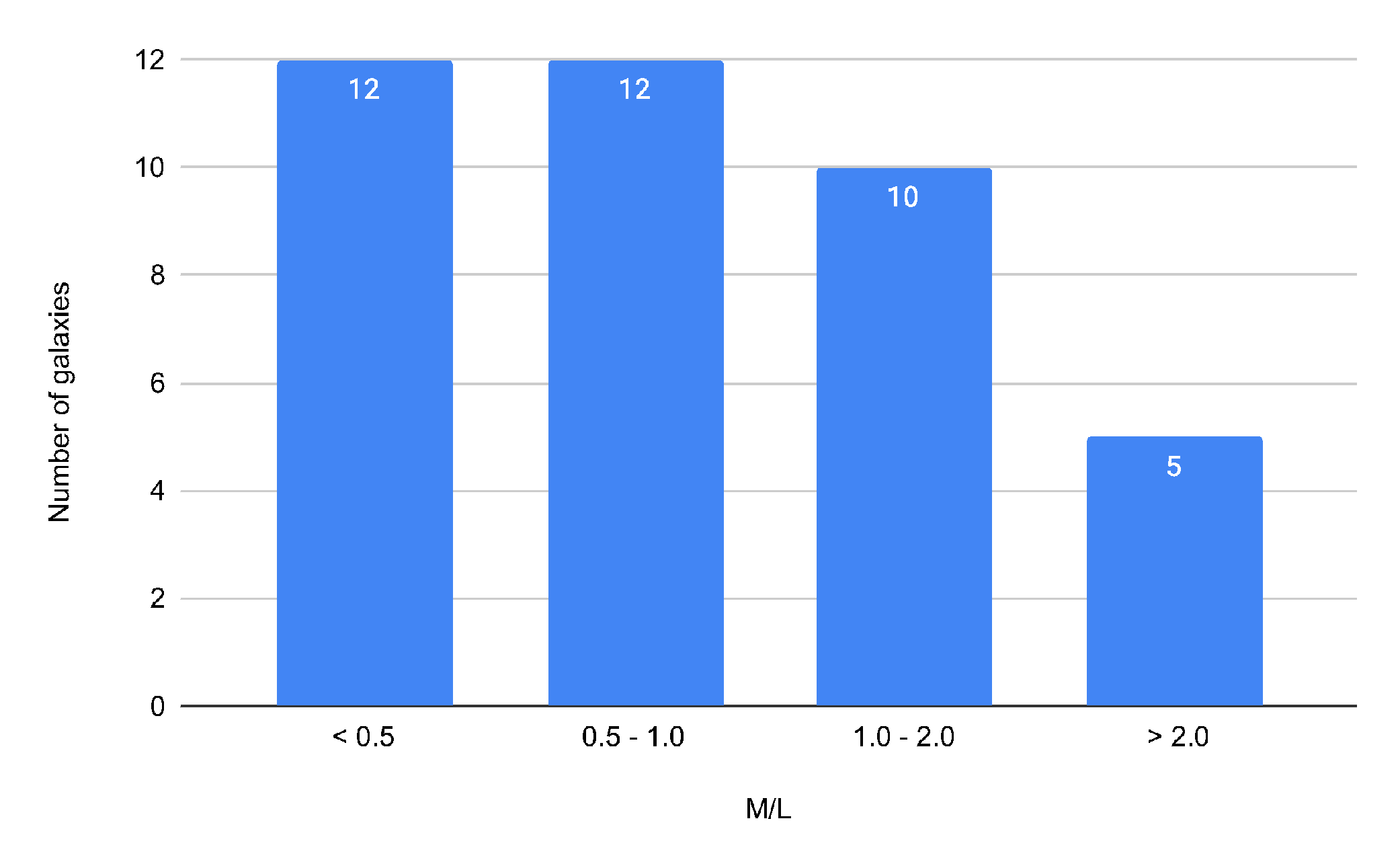}
\caption{ Distribution of the mass-to-light ratio values $\Upsilon_d$ for the disk  for the set of the 27 considered bulgeless galaxies with a fixed $R=16$ kpc.}
\label{fnML2}
\end{figure}
\end{center}

The distribution of the $\Upsilon_d$ values for $R=16$ kpc are represented in Fig.~\ref{fnML2}. There are no significant differences in the distribution of the mass-to-light ratio as compared to the varying $R$ case. For fixed $R=16$ kpc  the average value of $\Upsilon_d$ is $\left<\Upsilon_d\right>=0.916$, a value slightly smaller than the value obtained for varying $R$.

\begin{center}
\begin{figure}[h!]
\centering
\includegraphics[width=8.5cm]{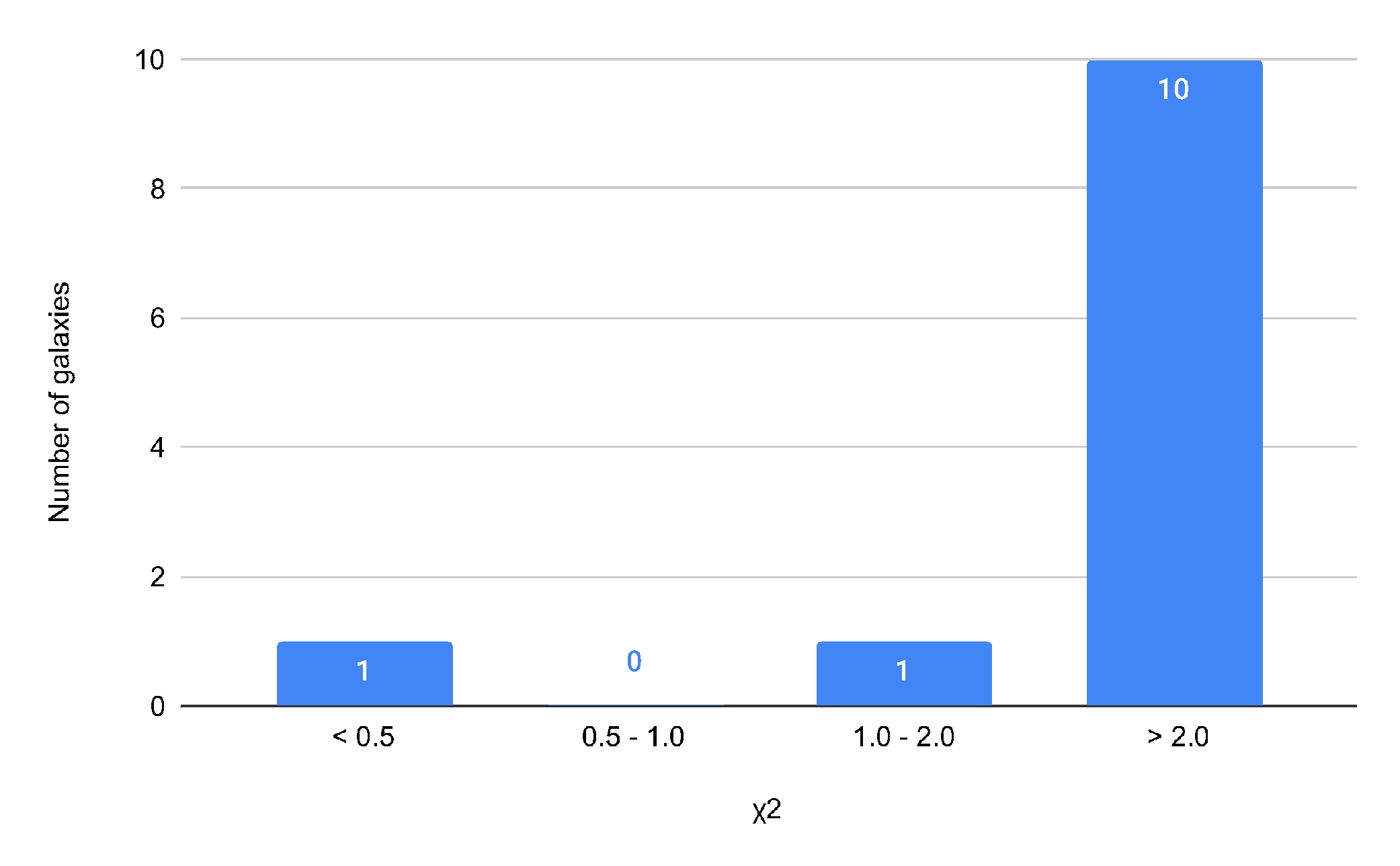}
\caption{The distribution of the $\chi^2$ values for a set of 12 galaxies for $R=16$ kpc, $\Upsilon _d=0.5$, and $\Upsilon _b=0$.}
\label{fn3}
\end{figure}
\end{center}

It is also interesting to investigate the effect of the $M/L$ ratio on the overall quality of the fitting. For this we have selected a set of 12 bulgeless galaxies, with $\Upsilon _b=0$, and we have performed again the fitting of the rotation curves. We present only the distribution of the $\chi^2$ values, which can be seen in Fig.~\ref{fn3}.

As one can see from Fig.~\ref{fn3}, fixing both $R$ and $\Upsilon _d$ induces a drastic change in the quality of the fit, with most of the galaxies having $\chi^2$ values greater than 2. A similar situation does appear in the case of galaxies with bulge, if one fixes $R=16$ kpc, and $\Upsilon _d=0.5$, respectively, and we let $\Upsilon _b$ to vary. In Fig.~\ref{fn4} we present the $chi^2$ distribution of the rotation curves fitting with fixed $R=16$ kpc, fixed $\Upsilon _d=0.5$, and varying $\Upsilon_b$ (left panel), and $\Upsilon_d=\Upsilon _b=0.5$ (right panel), respectively.

\begin{widetext}
 \begin{center}
\begin{figure*}[h!]
\centering
\includegraphics[width=8.5cm]{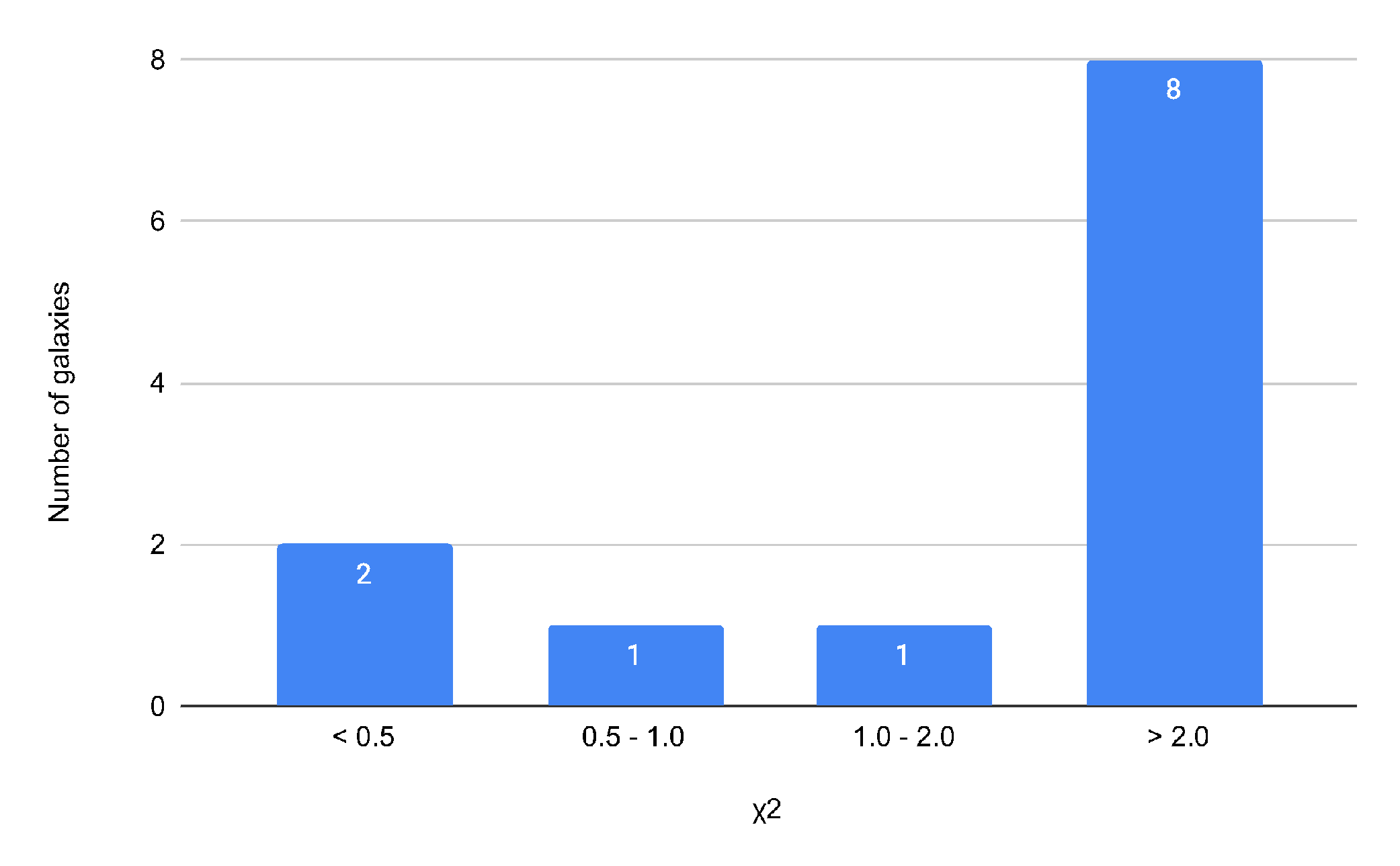}
\includegraphics[width=8.5cm]{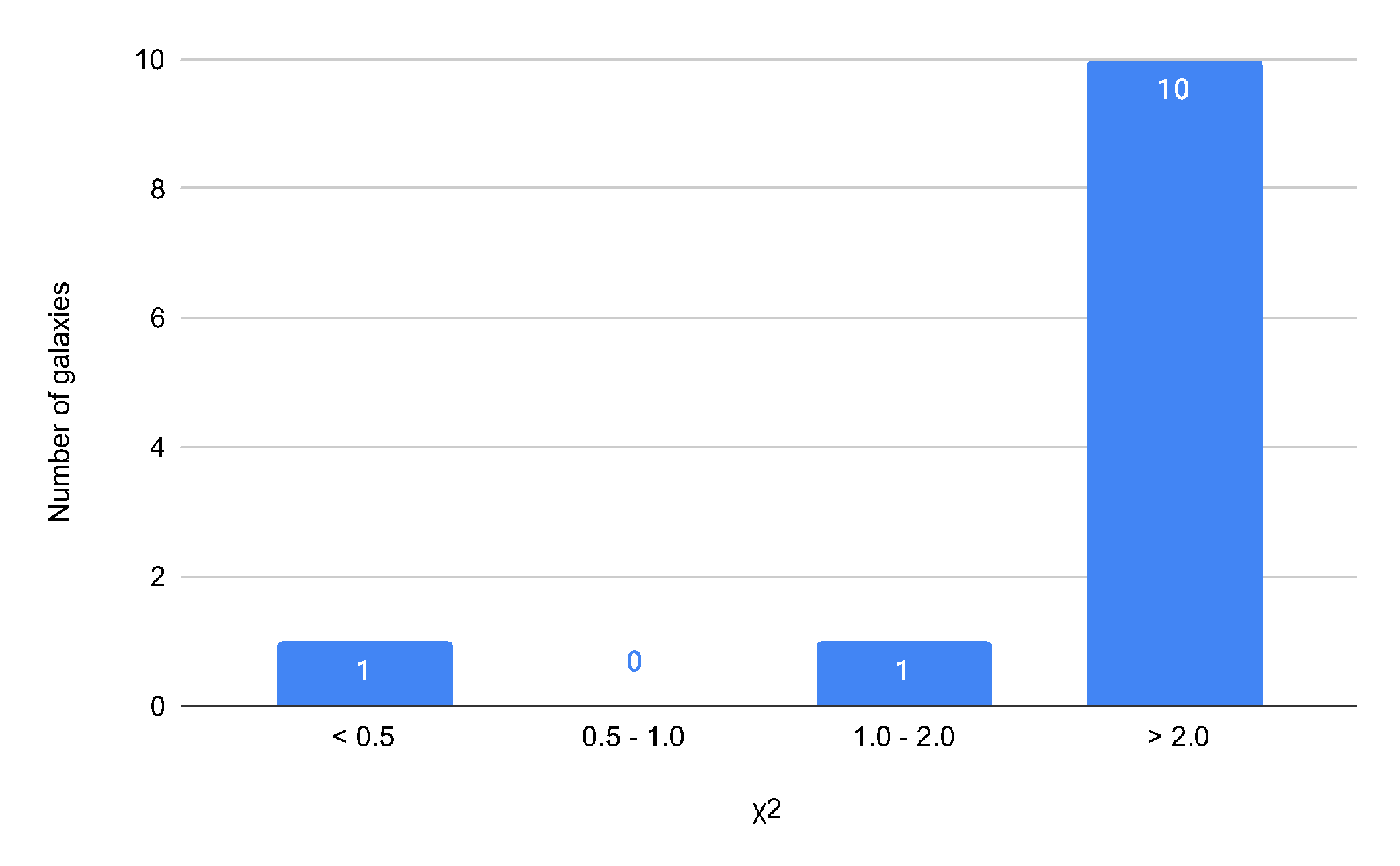}
\caption{Distribution of the $\chi^2$ values for a set of 12 galaxies with $R=16$ kpc, $\Upsilon _d=0.5$, and variable $\Upsilon _b$ (left panel), and $R=16$ kpc, and $\Upsilon _d=\Upsilon _b=0.5$ (right panel), respectively. }
\label{fn4}
\end{figure*}
\end{center}
\end{widetext}

Hence, the adopted numerical values for $M/L$ have a major effect on the quality of the fittings, and they may represent a major observational test for the Bose-Einstein Condensate models, once the range of values of $\Upsilon _b$ and $\Upsilon_d$ are strongly constrained by using astrophysical observations.

\subsection{Comparison of the Bose-Einstein Condensate dark matter models}

In the following we perform a comparative analysis for the theoretical models considered in the present work. Hence, we consider the simple static,nonrotating BEC model, in the absence of any random components, for which the tangential velocity is given by Eq.~(\ref{73}). We denote this model by SBEC. Secondly, we consider the contribution of the baryonic matter (BM) to this model, thus obtaining the model SBEC+BM. The Bose-Einstein Condensate dark matter model in the presence of random fluctuations is represented as RBEC, while in the presence of the baryonic matter, the model is labeled as RBEC+BM. In order to perform the comparison of these models we have selected a group of six galaxies from our sample, four of them without bulge velocity data, and two with bulge velocities. Even that this sample is small, it can be considered as representative for the set of galaxies we are considering in our study. The comparative behavior of the tangential velocity curves obtained by fitting the SPARC observational data for the six galaxies with the four models are represented in Fig.~\ref{C1}.   

\begin{widetext}
\begin{center}
\begin{figure*}[htbp]
\centering
\includegraphics[width=18cm]{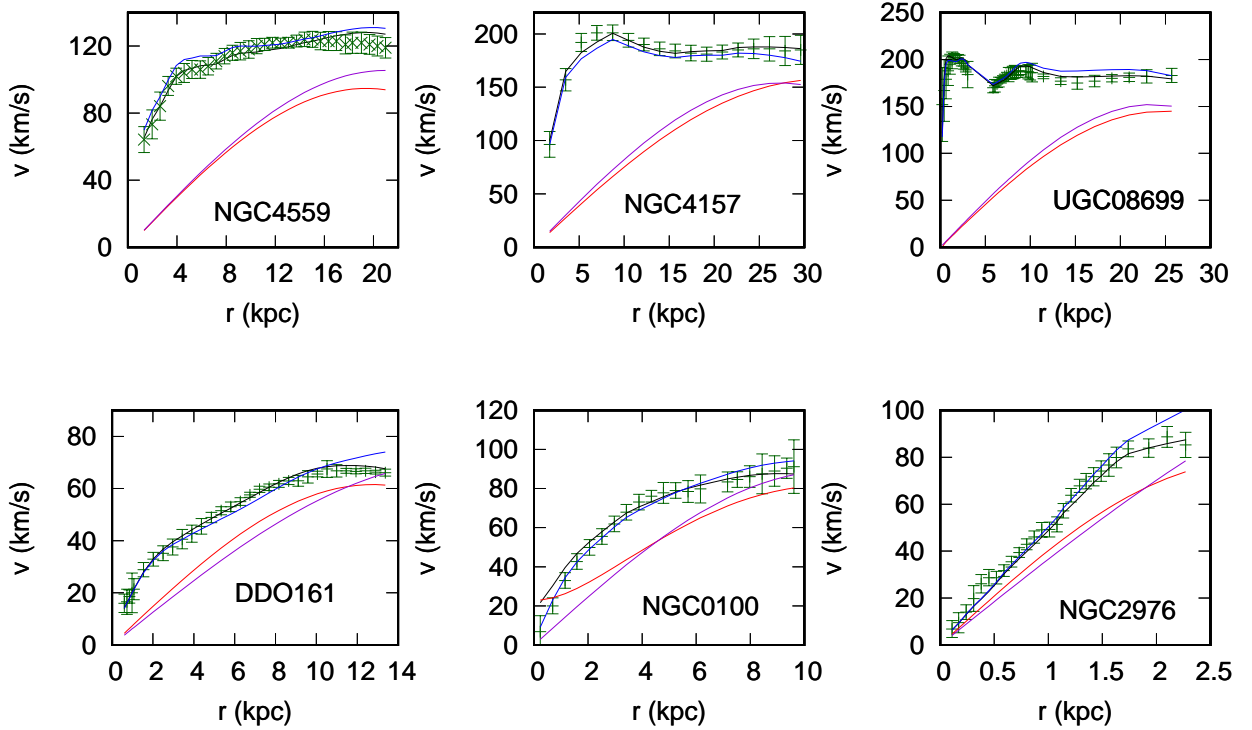}
\caption{Comparative analysis of four Bose-Einstein Condensate dark matter models for 6 galaxies. The different theoretical models are represented as follows:  RBEC - red curve, RBEC + BM - black curve, SBEC - dark violet curve, and SBEC + BM - blue curve, respectively. The observational data for the rotation curves $v_{obs}$ are represented in the dark green colour.}
\label{C1}
\end{figure*}
\end{center}
\end{widetext}

The results of the fitting of the four models are summarized in Table~\ref{C1}, which is presented as an Appendix.  Overall, the RBEC+BM model gives the best description of the observational data within the framework of the present formulation of the Bose-Einstein Condensate dark matter model. The quantitative differences between the predictions of the RBEC+BM and SBEC+BM models can be estimated, for different galaxies, and at different distances from the galactic center, as follows: for the galaxy DDO161, at $r = 13.2$ kpc, the differences $\Delta v$ between the predictions of the two models is $\Delta v=6.66$ km/s, while for $r=6$ kpc, the difference is $\Delta v=2.22$ km/s. For the galaxy  NGC0100 we obtain the values $\Delta v=7.14$ km/s at $r = 9.6$ kpc, and $\Delta v=4.28$ at $r=2$ kpc. For the considered sample of six galaxies the differences between the numerical predictions of the two models are in the range $\Delta v\in \left(4,13\right)$ km/s, and these differences are dependent on the distance from the galactic centre of the orbiting gas cloud. 

\section{Discussions and final remarks}\label{sect4}

If dark matter consists of bosonic particles, if these particles were produced in the early Universe, and if the laws of nature, as we know them from the Earth, are universal, then presently dark matter should be in the form of a Bose-Einstein Condensate, as a result of a cosmological transition that occurred when the temperature of the Universe became smaller than the critical transition temperature. The astrophysical and cosmological implications of the "pure" bosonic dark matter has been intensively investigated recently. In the present paper we have investigated the properties of the Bose-Einstein Condensate galactic dark matter halos, by extending the previous investigations in two directions, by pointing out the effects of the baryonic matter, and of the random gravitational effects on their structure and dynamical behavior.

Bose-Einstein Condensate dark matter models have been extensively used to model the galactic properties. It has been suggested that galaxies may form by the condensation of a scalar field in the very early Universe, initially forming Bose–Einstein Condensate drops, with the haloes of the galaxies being large structures made of a scalar field \cite{RM1}. If this is indeed the case, all galaxies must be very similar in their structure and global properties, like, for example, mass and radius. Fittings of the Bose-Einstein Condensate dark matter haloes with the high-resolution rotation curves of a sample of 13 low-surface-brightness galaxies were performed in \cite{RM1}, and a comparison with the fits obtained using Navarro–Frenk–White and pseudo-isothermal  profiles were also considered. A better agreement of the Bose-Einstein Condensate profiles with observations was found for this dark matter dominated sample of galaxies. However, the universality of the properties of the galactic dark matter halos may be lost, especially in the case of big galaxies \cite{RM2}. By considering that dark matter is a self-interacting real scalar field in a thermal bath at temperature $T$, by assuming an initial $Z_2$ symmetric potential, it turns out that as the universe expands, the temperature drops, with the $Z_2$ symmetry spontaneously broken. Hence, the scalar field reaches a new minimum, thus breaking the universal nature of the halo. An alternative approach to the Bose-Einstein Condensate dark matter was investigated in \cite{RM3}, by introducing several quantum states as a realistic model for a scalar field dark matter halo. By including these excited states one can reproduce the rotation curves of high-resolution low surface brightness  and SPARC galaxies by considering two scalar field dark matter profiles. The first is the soliton+Navarro-Frenk-White profile in the fuzzy dark matter model, obtained from the cosmological simulations of the dynamics of a real, non-interacting scalar field  at zero temperature. As for the second profile, the multistate scalar field dark matter profile was adopted, representing an exact solution to the Einstein–Klein–Gordon equations for a real, self-interacting scalar field, with finite temperature dependent potential. For the boson mass the range $0.212< m_{\psi}/\left(10^{-23}\;{\rm eV/c^2}\right) < 27.0$ was obtained.

In the present study we have investigated in detail the effects of the baryonic matter on the structure of the halo. A strong motivation to consider very carefully the role of the baryons is represented by the recent interesting results indicating the (unexpected) role the baryonc matter plays in the dark matter halos.
An explicit, and intriguing, empirical relation between the total acceleration of a particle in the galactic halo $a_{tot}(r)=v_{tot}^2(r)/r=GM_{tot}(r)/r^2$ and the acceleration of the baryonic matter $a_{b}(r)=v_{b}^2(r)/r=GM_b(r)/r^2$ was found in \cite{McG,Lel} in the form of the so-called radial acceleration relation,
\be\label{rar}
a_{tot}(r)=\frac{a_{b}(r)}{1-e^{-\sqrt{a_{b}(r)/a_+}}},
\ee
where $a_+=1.20\times 10^{-10}\;{\rm m/s^2}$. To fit the mean radial acceleration relation to 175 individual galaxies in the SPARC database the Markov chain Monte Carlo method was used in \cite{Lel1}, marginalizing over stellar mass-to-light ratio, galaxy distance, and disk inclination. For the vast majority of galaxies acceptable fits with astrophysically reasonable parameters were found, with the data consistent with the action of a single effective force law. Hence, the assumed universality of the acceleration scale, involving a tight relation between total acceleration and baryonic matter, if correct, is crucial for the understanding of dark matter, and galactic structure.

By assuming that the total mass of a galaxy can be represented as the sum of the dark and baryonic matter, $M_{tot}(r)=M_{DM}(r)+M_b(r)$, the radial acceleration relation (\ref{rar}) can be reformulated as a relation between the dark matter and baryonic masses,
\be
M_{DM}(r)=\frac{M_b(r)}{2}e^{-\sqrt{\frac{GM_b(r)}{4a_+r^2}}}{\rm csch}\left[\sqrt{\frac{GM_b(r)}{4a_+r^2}}\right],
\ee
where ${\rm csch}$ denotes the hyperbolic cosecant function ${\rm csch}(x)=1/\sinh (x)$. In the case of the static condensate the mass distribution of the dark matter is given by Eq.~(\ref{72}). Hence, in this case we obtain the following relation between the baryonic and dark matter masses,
\bea
&&\frac{M_b(r)}{2}e^{-\sqrt{\frac{GM_b(r)}{4a_+r^2}}}{\rm csch}\left[\sqrt{\frac{GM_b(r)}{4a_+r^2}}\right]=\nonumber\\
&&\frac{4\rho _cR^3}{\pi^2}\left[\sin \left(\frac{\pi r}{R}\right)-\frac{\pi r}{R}\cos\left(\frac{\pi r}{R}\right)\right].
\eea

In the low acceleration limit $a_b(r)<<a_+$, we obtain $a_{tot}=\sqrt{a_+a_b}$, and
\bea
&&\sqrt{\frac{GM_b(r)}{a_+r^2}}\left[1-\sqrt{\frac{GM_b(r)}{a_+r^2}}\right]\approx \nonumber\\
&&\frac{4\rho _cR^3}{\pi^2}\left[\sin \left(\frac{\pi r}{R}\right)-\frac{\pi r}{R}\cos\left(\frac{\pi r}{R}\right)\right].
\eea

In the limit $r\rightarrow R$, we obtain the following relation between the total condensate dark matter mass, and the baryonic mass,
 \be
 M_{DM}(R)=\frac{M_b(R)}{2}e^{-\sqrt{\frac{GM_b(R)}{4a_+R^2}}}{\rm csch}\left[\sqrt{\frac{GM_b(R)}{4a_+R^2}}\right].
 \ee

 In the present investigation, for the sake of mathematical simplicity, we have adopted for the baryonic matter density profile a simple exponential form, as given by Eq.~(\ref{49}). Hence the mass profile of the baryonic matter is given by
 \be
 M_b(r)=8\pi \rho _{bc}r_b^3\left\{1-\frac{1}{2r_b^2}e^{-r/r_b}\left[\left(r+r_b\right)^2+r_b^2\right]\right\},
 \ee
 which in the large $r$ limits tend to a constant, $M_b(r)\approx 8\pi \rho _{bc}r_b^3$, giving a uniform baryonic matter distribution. The constant $r_b$ is determined by the dimensionless parameter $\gamma$, via $r_b=R/( \pi \gamma)$. By taking into account the values of $\gamma$ obtained from the fitting of the rotation curves we obtain $r_b\approx 3$ kpc.

 The consequences of the constancy of $\mu_{DM}=\rho _sr_s$ was investigated for the ultralight scalar field dark matter model in \cite{RM4}. It was found that for this dark matter model $\mu _{DM}=648M_{\odot}/{\rm pc}^2$, and that the  Wave Dark Matter soliton profile is a universal feature of the dark matter halos.
 
 The exponential disk model is largely used in the modelling of galactic baryonic matter \cite{Ge1,Ge2}. However, it was found that for the SPARC data the exponential disk model underestimates the luminosity of the inner region for some galaxies, while for other data sets it under- or overestimates the luminosity of the outer region \cite{Ge1,Ge2}. In order to obtain a better description of the baryonic matter properties one could use, for example, the Tempel–Tenjes model, in which the spatial
luminosity density distribution of each visible component in the galaxy is given by $l(a)=l(0)\exp \left[-\left(a/ka_0\right)^{1/N}\right]$, where $a=\sqrt{r^2+z^2/q^{2}}$, where $r$ and $z$ are cylindrical coordinates, $q$ is the axis ratio, $l(0)=hL\left(4\pi q a_0^3\right)^{-1}$, $a_0$ is the harmonic mean radius of the considered component, and $k$ and $h$ are scaling parameters, respectively. In a two-component baryonic model the spatial mass density can be constructed as $\rho (a)=\Upsilon _bl_b(a)+ \Upsilon _d l_d(a)$ \cite{Ge2}, where  $l_b(a)$ and $l_d(a)$ denote the spatial luminosity densities of the bulge and disk, respectively, and $\Upsilon _b$ and $\Upsilon _d$ are the corresponding  $(M/L)$ ratios \cite{Ge2}. However, despite its limitations, the simple exponential disk model used in the present work can still provide, at least on a qualitative level, a first order approximate description of the complicated baryonic structure and distribution in galaxies.

 For the  random potential we have adopted a Gaussian distribution, which is well motivated physically, and it also allows obtaining the general solution of the basic density equation in an exact analytic form. The random potential is determined by two parameters $(\sigma, \theta)$, and it has a significant effect on the overall structure of the condensate. The effective random density contribution to the total energy is of the same order of magnitude as the gravitational energy of the galaxy, including the baryonic matter contribution, indicating the presence of strong disorder effects in the considered set of galaxies. One possible source of disorder may be generated by the transfer angular momentum and energy from the
galactic baryonic matter to the dark matter halo, which induces a random dynamics in the system. Tidal interactions
caused by distant external sources can also be modelled with the help of random potentials. Hence in the complicated galactic background dark matter halos are exposed to a large number of gravitational effects from nearby galaxies or stellar encounters. Such essentially random processes must be taken into account when considering condensate dark matter structure, and dynamics.

All the Bose-Einstein Condensate dark matter models assume the existence of a universal constant $R=\sqrt{a\hbar ^2/Gm^3}$, determined by the mass of the dark matter particle, its scattering length, and two fundamental constants, $\hbar$ and $G$. From a physical point of view $R$ can be interpreted as the radius of the static condensate, in the presence of the quartic confining potential only. Checking the constancy of this parameter is crucial for the astrophysical viability of the condensate dark matter models. However, standard fitting of the galactic rotation curves usually provides a large range of values for $R$, with large dispersion measures. For example, for the 27 bulgeless SPARC galaxies, with a few exceptions $R$ is in the range $R\in (11,20)$ kpc, while for galaxies with a bulge $R\in (20,50)$ kpc. These results raise the problem of the compatibility of the Bose-Einstein Condensate model with observations, and of the equivalence of the condensate dark matter structures in galaxies with and without a bulge. To investigate this problem we have adopted an average value for $R$, and we have redone the fittings. As one can see from  Tables~\ref{tableg3a} and \ref{tableg4}, there are no significant changes in the values of $\chi ^2$ as compared to the case in which $R$ is allowed to vary freely. These numerical results strongly indicate that by taking into account the presence of various physical processes that occur in realistic galactic systems the constancy of $R$ can be achieved without significant effects on the statistical analysis, and $\chi^2$ values.
For the adopted value of $R$ we obtain for the mass $m$ of the dark matter particles the expression
\be
m =\left( \frac{\pi ^{2}\hbar ^{2}a}{GR^{2}}\right) ^{1/3}\approx
0.91\times 10^{-2}\times \left[ a\left( \mathrm{fm}\right) \right] ^{1/3}\;\mathrm{eV}.
\ee
Hence the mass of the dark matter particle is fully determined by its scattering length. Ultra-light dark matter particles with masses of the order of $m\approx 10^{-22}$ eV would require scattering lengths of the order of $a\approx 10^{-60}$ fm, a value which is theoretically possible if we take into account the extremely weak interaction between dark matter particles. On the other hand, a scattering length of $a\approx 1$ fm would give a mass $m\approx 10^{-2}$ eV, while a scattering length of $a\approx 10^6$ fm would bring the mass of the dark matter particle close to the value of $m\approx 1$ eV, representing the absolute mass scale of neutrinos \cite{Aker}. However, until more information on the physical nature of dark matter would become available, the exact value of the mass of the dark matter particle will remain elusive. Astrophysical and astronomical data alone cannot give a final answer to this problem.

The Bose-Einstein Condensate dark matter model represents an attractive alternative to the standard cold dark matter model. Moreover, it is an astrophysical realization of a well known process that can be studied and tested in the laboratory. In the present paper we have introduced some basic, but realistic features of the Bose-Einstein Condensate dark matter model that can help in its observational testing.

\section*{Acknowledgments}

We would like to thank the anonymous reviewer for comments and suggestions that helped us to significantly improve our work. We  thank Prof. Thomas Nattermann for useful suggestions and help during the preparation of the manuscript. The work of TH is supported by a grant of the Romanian Ministry of Education and Research, CNCS-UEFISCDI, project number PN-III-P4-ID-PCE-2020-2255 (PNCDI III).

\appendix

\section{Results of the comparative fitting of four BEC dark matter models}

The results of the fitting of the four considered Bose-Einstein Condensate dark matter models (SBEC, SBEC+BM, RBEC, RBEC+BM) are presented in Table~\ref{C1}. Each model is characterized by a different set of parameters, with the simplest model, the SBEC model, described by only two independent physical parameters, $\rho_c$ and $R$, respectively. All the values of the parameters have been obtained by fitting the models with the observational data of the SPARC database. For the model comparison from the total set of 39 galaxies considered in the present study we have selected a set of six galaxies only, two of them also having bulge velocity data, while for four galxies no such data are provided.      

\newpage

 \begin{widetext}
 \begin{center}
\begin{table}[htbp]
\centering
\begin{tabular}{|c|c|c|c|c|c|c|c|c|}
\hline
RBEC+BM && &&&&&& \\
\hline
${\rm Galaxy}$ & $R \;({\rm kpc}) $ & ${\rho}_{c} \left(10^{-24} {\rm g/cm}^{3}\right)$ & ${\rho}_{bc} \left(10^{-24} {\rm g/cm}^{3}\right)$ & $\omega \;(10^{-16} {\rm s}^{-1})$ & $\sigma $ & $ \theta $ & $ \gamma $ & $ \chi^{2} $ \\
\hline
DDO161 & 14.353 & 0.2060 & 0.015 & 0.4070 & 0.0123 & 0.00139 & 1.9504 & 0.293  \\
\hline
NGC0100 & 10.69 & 0.6080 & 0.1216 & 0.9272 & 0.011 & 0.000127 & 1.7535 & 0.225 \\
\hline
NGC2976 & 3.18 & 6.681 & 1.2762 & 1.7040 & 0.029 & 0.00350 & 2.108 & 0.389 \\
\hline
NGC4559 & 21.930 & 0.251 & 0.0502 & 0.476 & 0.0265 & 0.00112 & 1.692 & 0.135 \\
\hline
NGC4157 & 30.45 & 0.243 & 0.0486 & 1.215 & 0.0160 & 0.0022 & 1.97 & 0.208 \\
\hline
UGC08699 & 26.48 & 0.315 & 0.065 & 1.149 & 0.0136 & 0.00185 & 1.9552 & 0.284  \\
\hline
\end{tabular}
\begin{tabular}{|c|c|c|c|c|c|c|c|c|}
\hline
RBEC & &&&&&&&\\
\hline
${\rm Galaxy}$ & $R \;({\rm kpc})$ & ${\rho}_{c} \left(10^{-24} {\rm g/cm}^{3}\right)$ & ${\rho}_{bc} \left(10^{-24} {\rm g/cm}^{3}\right)$ & $\omega \;(10^{-16} {\rm s}^{-1})$ & $\sigma $ & $ \theta $ & $ \gamma $ & $ \chi^{2} $ \\
\hline
DDO161 & 14.413 & 0.211 & 0.0161 & 0.412 & 0.0142 & 0.00128 & 1.8309 & 0.538  \\
\hline
NGC0100 & 12.380 & 0.507 & 0.0924 & 0.931 & 0.0130 & 0.00150 & 1.1530 & 0.665 \\
\hline
NGC2976 & 3.270 & 6.757 & 0.8834 & 1.718 & 0.0260 & 0.00320 & 2.1160 & 0.819 \\
\hline
NGC4559 & 21.930 & 0.218 & 0.0302 & 0.482 & 0.0283 & 0.00126 & 1.7140 & 0.825 \\
\hline
NGC4157 & 31.790 & 0.235 & 0.0423 & 1.234 & 0.0180 & 0.00250 & 1.8300 & 1.016 \\
\hline
UGC08699 & 26.270 & 0.327 & 0.0870 & 1.162 & 0.0131 & 0.00173 & 1.8420 & 0.123  \\
\hline
\end{tabular}
\begin{tabular}{|c|c|c|c|c|c|}
\hline
SBEC+BM &&&&&\\
\hline
${\rm Galaxy}$ & $R \;({\rm kpc})$ & ${\rho}_{c} \left(10^{-24} {\rm g/cm}^{3}\right)$ & $ \Upsilon_{d} $ & $ \Upsilon_{b} $ & $ \chi^{2} $ \\
\hline
DDO161 & 19.300 & 0.160 & 0.8600 & & 0.337 \\
\hline
NGC0100 & 11.690 & 0.614 & 0.7940 & &0.219\\
\hline
NGC2976 & 4.880 & 6.981 & 0.2802 & &0.403 \\
\hline
NGC4559 & 24.410 & 0.214 & 0.8200 & & 0.387 \\
\hline
NGC4157 & 30.970 & 0.257 & 0.5100 & 0.0810 & 0.318 \\
\hline
UGC08699 & 26.930 & 0.364 & 1.3670 & 0.5140 & 0.423 \\
\hline
\end{tabular}
\begin{tabular}{|c|c|c|c|}
\hline
SBEC &&&\\
\hline
${\rm Galaxy}$ & $R \;({\rm kpc})$ &${\rho}_{c} \left(10^{-24} {\rm g/cm}^{3}\right)$ & $ \chi^{2} $ \\
\hline
DDO161 & 19.600 & 0.150 & 0.429 \\
\hline
NGC0100 & 12.570 & 0.584 & 0.678\\
\hline
NGC2976 & 6.150 & 5.174 & 0.823 \\
\hline
NGC4559 & 23.910 & 0.227 & 0.932 \\
\hline
NGC4157 & 31.280 & 0.283 & 1.023 \\
\hline
UGC08699 & 26.980 & 0.371 & 0.132 \\
\hline
\end{tabular}
\caption{Fitting results of the SBEC, SBEC+BM, RBEC, and RBEC+BM models with the SPARC observational data.}\label{C1}
\end{table}
\end{center}
\end{widetext}

\end{document}